\documentclass[aps,prb,longbibliography,twocolumn,floatfix,superscriptaddress,nofootinbib]{revtex4-2}

\usepackage{graphicx}
\usepackage{amsmath}
\usepackage{braket}
\usepackage{amssymb}
\usepackage{mathtools}
\usepackage{makeidx}
\usepackage{amsfonts}
\usepackage{bm}
\usepackage{tikz}
\usepackage{xcolor} % for RGB colors
\usepackage[colorlinks=true, linkcolor=blue, citecolor=blue, urlcolor=blue]{hyperref}
\usetikzlibrary{shapes,shadows}

\usepackage{soul} % for strikethrough \st{text}

\definecolor{linkcolour}{rgb}{0,0,1}
\newcommand{\flc}[1]{\textcolor{linkcolour}{(#1)}}

\definecolor{crimson}{rgb}{0.86,0.08,0.24}

\definecolor{minorcolor}{rgb}{0.00,0.70,0.22}

\definecolor{doubtcolor}{rgb}{0.6,0.3,0.6}

\begin{document}

% \title{\major{Entanglement Entropy and Lyapunov Exponents Reveal Topological Phase Transitions in Disordered SSH Chains}}
% \title{Probing Topological Phase Transitions in Quasiperiodic and Binary-Disordered SSH Chains via Entanglement Entropy and Lyapunov Exponents}
\title{Entanglement entropy as a probe of topological phase transitions}

\author{Manish Kumar}
\affiliation{Department of Physics, Indian Institute of Science Education and Research, Bhopal, Madhya Pradesh 462066, India}
\author{Bharadwaj Vedula}
\affiliation{Department of Physics, Indian Institute of Science Education and Research, Bhopal, Madhya Pradesh 462066, India}
\author{Suhas Gangadharaiah}
\affiliation{Department of Physics, Indian Institute of Science Education and Research, Bhopal, Madhya Pradesh 462066, India}
\author{Auditya Sharma}
% \email{auditya@iiserb.ac.in}
\affiliation{Department of Physics, Indian Institute of Science Education and Research, Bhopal, Madhya Pradesh 462066, India}
\date{\today}

\begin{abstract}
  Entanglement entropy (EE) provides a powerful probe of quantum
  phases, yet its role in identifying topological phase transitions in
  disordered systems remains underexplored. We introduce an exact
  EE-based framework that captures topological phase transitions even
  in the presence of disorder.  Specifically, for a class of
  Su-Schrieffer-Heeger (SSH) model variants, we show that the
  difference in EE between half-filled and near-half-filled ground
  states, $\Delta S^{\mathcal{A}}$, vanishes in the topological phase
  but remains finite in the trivial phase, a direct consequence of
  edge-state localization. This behavior persists even in the presence
  of quasiperiodic or binary disorder. By analyzing domain-wall
  configurations in the SSH chain, we further show how subsystem
  tuning allows one to distinguish genuine topological zero-energy
  eigenstates from trivial localized states. Exact phase boundaries,
  derived from Lyapunov exponents via transfer matrices, agree closely
  with numerical results from $\Delta S^{\mathcal{A}}$ and the
  topological invariant $\mathcal{Q}$, with instances where
  $\Delta S^{\mathcal{A}}$ outperforms $\mathcal{Q}$. Our results
  highlight EE as a robust diagnostic tool and a potential bridge
  between quantum information and condensed matter approaches to
  topological matter.
\end{abstract}

\maketitle

\section{Introduction}
\label{sec:introduction}

Entanglement entropy (EE) has become a
central tool in the study of quantum many-body systems, and quantum
phase transitions (QPTs)~\cite{Osterloh2002, PhysRevLett.90.227902,
  PhysRevA.66.032110, RevModPhys.82.277}. Defined as the von Neumann
entropy of a subsystem, it quantifies
the degree of quantum correlations between parts of a
system~\cite{PhysRevA.53.2046}. In conventional systems, QPTs are
typically associated with symmetry breaking
phenomena~\cite{Landau_1937, SciPostPhysLectNotes,
  landau1980condensed}. However, many exotic phases such as
topological insulators and fractional quantum Hall states, do not fit
within this paradigm. Topological phases of matter have been an active
field of research in condensed matter
physics~\cite{RevModPhys.82.3045, RevModPhys.83.1057, Moore2010,
  PhysRevB.76.045302, Bernevig2013, Asb_th_2016} ever since the
discovery of the quantum Hall effect~\cite{PhysRevLett.45.494}. These
phases are distinguished by topological
invariants~\cite{PhysRevB.75.121306, PhysRevLett.97.036808,
  PhysRevB.75.121403, PhysRevLett.95.146802,
  PhysRevB.74.195312,PhysRevB.79.195321} that remain robust against
local perturbations~\cite{PhysRevB.80.125327, PhysRevB.85.205136,
  PhysRevX.2.031008, PhysRevB.99.035146}. The far-reaching
implications of these discoveries~\cite{PhysRevB.92.085118,
  RevModPhys.88.035005, RevModPhys.89.041004, Wieder2022}, have
inspired work across fields including photonic
systems~\cite{CRPHYS_2016__17_8_808_0, Lu2014, RevModPhys.91.015006},
ultracold atoms~\cite{ZHANG2018253, Galitski2013,
  RevModPhys.91.015005, Goldman2016}, and quantum information
technologies~\cite{PhysRevLett.91.147902, PAWLAK20191}.

Traditionally, topological phase transitions are identified using
momentum-space invariants like the winding
number~\cite{Li_2015,Malakar_2023}, and the $\mathcal{Z}_2$ invariant~\cite{PhysRevResearch.3.L012001} which rely on translational
symmetry and thus break down in disordered systems. To address this,
real-space alternatives, such as real space winding numbers
~\cite{PhysRevLett.113.046802} and scattering coefficients sensitive
to edge-state localization~\cite{PhysRevB.83.155429, Zhang_2016}, have
been developed.  Among these, EE has emerged as a uniquely powerful
probe. Its non-local character~\cite{PhysRevB.101.195117} enables
detection of topological properties even in many-body
states~\cite{PhysRevB.98.045120,zeng2018} where conventional order
parameters fail.  In this work we provide an exact framework to use
many body EE to study topological phase transitions which works even
in the presence of disorder.

Despite many studies, a comprehensive understanding of EE across
topological phase transitions remains incomplete. While EE's
sensitivity to distinct topological phases is
well-established~\cite{PhysRevB.101.195117,PhysRevB.98.045120,
  PhysRevB.77.155111, PhysRevB.89.094512, Jiang2012,
  PhysRevB.101.014301, PhysRevB.77.054433, PhysRevB.84.195120,
  PhysRevB.83.075102, PhysRevB.81.064439, zeng2018, PhysRevLett.104.130502}, the microscopic
connection between entanglement patterns and edge-state localization,
especially in disordered systems, remains poorly understood. This gap
motivates our work, where we develop a systematic framework using many
particle EE to distinguish topological and trivial phases.  For this
purpose, we choose the Su-Schrieffer-Heeger (SSH)
model~\cite{PhysRevLett.42.1698, PhysRevB.22.2099}, which was first
introduced to describe both elastic and electronic properties of
polyacetylene, a quasi-one-dimensional polymer chain. It is a 1D
tight-binding model with non-interacting spinless fermions, and is a
standard test-bed for explorations on topological systems. One of the
most striking features of the SSH chain is the presence of chiral
symmetry protected zero-energy edge states in the topologically
non-trivial phase which are robust to local perturbations that
preserve the bulk gap and symmetry~\citep{PhysRevB.89.085111,
  quasi_SSH, Diptiman_sen, binary_SSH, PhysRevB.109.195124,
  PhysRevA.105.063327, PhysRevResearch.6.L042038, Li_2021,
  PhysRevB.107.035113, Xu2024}.

\begin{figure}
    \centering
    \resizebox{0.48\textwidth}{!}{  % Rescale TikZ to desired width
    \begin{tikzpicture}
        % Define colors
        \definecolor{bcolor}{RGB}{185,56,124}  % burgundy
        \definecolor{bgcolor}{RGB}{185,56,124}      % Dashed line color
        \definecolor{acolor}{RGB}{30,144,255}       % DodgerBlue for A
%        \definecolor{bcolor}{RGB}{255,140,0}        % DarkOrange for B
        \definecolor{atext}{RGB}{255,255,255}       % White text on A
        \definecolor{btext}{RGB}{0,0,0}             % Black text on B
        \definecolor{darkgreen}{RGB}{0,100,0}
        % Parameters
        \def\Alevel{0}
        \def\Blevel{0.7}

        % Draw 3D-looking sticks (fake cylinders) connecting A to B
\foreach \x in {0,2,...,10} {
    \shade[left color=black!80, right color=black!40, shading=axis]
        (\x,\Alevel) -- ({\x+1},\Blevel) -- ++(0.1,0.1) -- ({\x+0.1},\Alevel+0.1) -- cycle;
}
\foreach \x in {1,3,...,11} {
    \shade[left color=black!80, right color=black!40, shading=axis]
        (\x,\Blevel) -- ({\x+1},\Alevel) -- ++(0.1,0.1) -- ({\x+0.1},\Blevel+0.1) -- cycle;
}

        % Draw the A and B markers as 3D spheres
\foreach \x in {0,1,...,12} {
    \ifodd\x
        % B marker (higher)
        \shade[ball color=bcolor, shading=ball] (\x,\Blevel) circle(8pt);
        \node[text=atext] at (\x,\Blevel) {\textbf{B}};
    \else
        % A marker (lower)
        \shade[ball color=acolor, shading=ball] (\x,\Alevel) circle(8pt);
        \node[text=atext] at (\x,\Alevel) {\textbf{A}};
    \fi
}

        % Draw dashed rectangle around markers 5–8
        \draw[darkgreen, dashed, line width = 3, rounded corners=4pt]
            (4.6,-0.4) rectangle (8.4,1.1);  % (x_min, y_min) to (x_max, y_max)
         
         % Add \mathcal{A} label above the green box
        \node[red] at (6.5, 1.5) {\huge $\mathcal{A}$};

    \end{tikzpicture}
    }
    \caption{SSH chain where the green dashed box shows subsystem
      $\mathcal{A}$ composed of a few unit cells deep inside the
      bulk. In this paper, the subsystems are denoted by calligraphic
      letters $\mathcal{A}$ and $\mathcal{B}$, while the lattice sites
      within each unit cell are labeled by $A$ and $B$.}
    \label{fig:subsystem_A}
\end{figure}

In this paper, we propose an EE based approach to study topological
phase transitions and demonstrate its robustness by performing an
independent transfer matrix based analysis. Specifically, we
analytically derive the topological-trivial phase boundaries by
calculating the Lyapunov exponent (LE) of edge states in the
topological region.  While this transfer-matrix-based calculation of
the LE has been widely used in the studies of mobility edges in
disordered systems~\cite{PhysRevLett.125.196604, Cai_2023,
  PhysRevLett.131.176401, PhysRevB.108.174202, PhysRevB.103.174205}
and Majorana fermionic systems~\cite{PhysRevB.83.155429, Zhang_2016,
  PhysRevLett.106.057001}, it has not been used in the context of the
SSH model. We demonstrate that the phase boundaries obtained from the
two methods match perfectly with each other.  We further use the
topological quantum number $\mathcal{Q}$~\cite{PhysRevB.83.155429}
which is a well-established quantity that characterizes the presence
or absence of Majorana bound states in 1D p-wave superconductors. We
also show an example with disorder where EE works better than
$\mathcal{Q}$ which exhibits some ambiguity. Finally, we present a
detailed analysis of domain-wall configurations in the SSH chain to
clarify the scope and interpretation of our entanglement-based
diagnostic. We show that while the EE-based quantity primarily detects
edge localization, combining it with a systematic tuning of the
subsystem allows one to distinguish genuine topological zero-energy
states from trivial localized states.  This domain-wall analysis
establishes a consistent framework for identifying topological phases
beyond simple localization criteria.

The paper is organized as follows. Section~\ref{sec:Model} introduces
the SSH model and its disordered variants, establishing the
theoretical framework for our study. In Section~\ref{sec:analytical},
we derive exact phase boundaries through the transfer matrix method,
computing the Lyapunov exponent to distinguish topological and trivial
phases. Section~\ref{entagnlement} demonstrates how the entanglement
entropy difference between half-filled and half-plus-one-filled
regimes serves as a robust numerical probe of the phase
transition. The physical origin of this behavior is elucidated in
Section~\ref{Occupation} through occupation number analysis, revealing
the edge localization of added particles in the topological
phase. Section~\ref{Casual_localized states} addresses the scenario of
topologically trivial states localized at the edges and clarifies the
conditions under which our entanglement-entropy-based diagnostic
distinguishes genuine topological edge states from accidental
edge-localized states. Section~\ref{Quantum number} discusses the
topological quantum number $\mathcal{Q}$ and its performance as a
diagnostic in relation to the other measures proposed here. In
Section~\ref{Domain_wall_section}, we further analyze domain-wall
configurations in the SSH chain and show how a systematic tuning of
the subsystem enables a clear distinction between topological and
non-topological localized states. Finally, Section~\ref{sec:Summary}
summarizes our results and discusses their broader
implications. Appendix~\ref{Appendix} presents additional results for
other parameter choices of binary disorder, demonstrating the
generality of our analytical framework.

\section{Models}
\label{sec:Model}
The Su--Schrieffer--Heeger (SSH) model~\cite{PhysRevLett.42.1698,
  PhysRevB.22.2099} describes a one-dimensional dimerized lattice with
two sites per unit cell and staggered nearest-neighbour hopping. In
second-quantized form the Hamiltonian is
\begin{equation}
H = \sum_{n=1}^L\left[t_1a_{n}^{\dagger}b_{n} + t_2 a_{n+1}^{\dagger}b_{n} + \text{H.c.}\right],
\label{Ham}
\end{equation}
where $a_{n}^{\dagger}$ ($b_{n}^{\dagger}$) creates a particle on
sublattice $A$ ($B$) of the $n^{\text{th}}$ unit cell, and $L$ is
the total number of unit cells. Since each unit cell contains two
sites, the total number of sites is $N=2 L$. The first term represents
intracell hopping with strength $t_1$, whereas the second corresponds
to intercell hopping with strength $t_2$. The competition between
these two hopping parameters determines the topological phase of the
system. In particular if $ |t_1| > |t_2| $ the system is in the
trivial phase while for $ |t_1| < |t_2| $, the system is in the
topological phase. The edge modes in the topological phase are
protected by chiral symmetry and remain stable against perturbations
that preserve this symmetry and do not close the bulk energy gap. In
this paper, apart from the clean SSH model, we extend the study to two
disordered but chiral symmetry preserving variants of the SSH model:
one in which the hopping terms are quasiperiodic and the other with
binary random disorder. Our results show that the proposed analytical
as well as numerical methods reliably detect topological phase
transitions even in the presence of disorder, demonstrating its
robustness and consistency with established results in the literature.

\subsection{Clean SSH model}

A convenient representation of the clean SSH model is to parameterize
it with average hopping amplitude $t$, and dimerization strength
$\lambda$. Thus the intracell and intercell hopping terms,
respectively, are given by
\begin{align}
  t_1 &= t-\lambda, \nonumber \\
  t_2 &= t+\lambda.
  \label{Clean_Ham}
\end{align}
Here we focus on the case $ t > 0 $; the results for $ t < 0 $ can
be obtained analogously. When $\lambda>0$ the intercell hopping
dominates, yielding a nonzero winding
number~\cite{Li_2015,Malakar_2023}, and, under open boundary
conditions, a pair of zero-energy edge states---signatures of a
topologically non-trivial phase.  For $\lambda<0$ the intracell
hopping prevails, the winding number vanishes, and no edge states
appear, so the phase is topologically trivial.  Hence the relative
strength of intercell and intracell hopping controls the topological
character of the SSH chain.

\subsection{SSH model with quasiperiodic disorder hopping}
Next, we introduce a quasiperiodic modulation in the hopping
amplitudes of the SSH model~\cite{quasi_SSH, Diptiman_sen}. The
modified intracell and intercell hopping terms, respectively, are
\begin{align}
  t_1&=t-\lambda-\delta \cos(2\pi\beta n+\phi), \nonumber \\
  t_2&=t+\lambda+\delta \cos(2\pi\beta n+\phi).
  \label{quasi_hopping_strength}
\end{align}
Note that this modification leaves the chiral symmetry intact. Here we
focus on the case with $\lambda>0$ and $\delta>0$. The parameter
$\beta = (\sqrt{5} - 1)/2$ (the inverse golden ratio) ensures
incommensurate modulation, preventing periodic recurrence and thus
preserving quasiperiodicity. Here, $\phi$ is an arbitrary global phase
drawn randomly from a uniform distribution in $[0,2\pi]$, ensuring
that the quasiperiodic potential explores all phase configurations
ergodically.  In previous studies, it has been conjectured that on
adding this quasiperiodic potential in the hopping term, the
topological phase transition occurs at
$\delta=t+\lambda$~\cite{quasi_SSH}. However, an exact derivation of
this phase boundary has been lacking. In this work, we analytically
derive the boundary using the Lyapunov exponent of the edge state
obtained via the transfer matrix method. Furthermore, we verify this
analytical result numerically using the entanglement entropy. Our
results show that for $\delta<t+\lambda$ the system behaves as a
topological insulator whereas for $\delta>t+\lambda$ it is a trivial
insulator.

\subsection{SSH model with random binary disorder hopping}
We introduce random binary disorder in the hopping strength in the SSH
model. To do this, we set the hopping strengths to be
\begin{align}
  t_1&=t-\lambda-\Delta_n, \nonumber \\
  t_2&=t+\lambda+\Delta_n.
  \label{random_hopping_strength}
\end{align}
Here, the $\Delta_n$ are drawn from a bimodal probability distribution function~\cite{binary_SSH}
\begin{equation}
\mathcal{P}(\Delta_n) = P \, \delta(\Delta_n - V_0) + (1 - P) \, \delta(\Delta_n - V_0 - W),
\label{Delta_value_distribution}
\end{equation}
where $W$ is the disorder strength and $V_0$ is the offset, which we fix at
$2$. Hence, $\Delta_n$ can
take values
\begin{equation}
\Delta_n =
\begin{cases}
2, & \text{with probability } P; \\
2 - W, & \text{with probability } 1 - P.
\end{cases}
\label{Delta_value}
\end{equation}
Although random binary disorder in the hopping terms of the SSH model
has been explored in earlier studies~\cite{binary_SSH}, the specific
form considered in Eq. \eqref{random_hopping_strength} has not been
previously analyzed.  In this work, we demonstrate that the LE
calculated via the transfer matrix method provides a reliable
analytical approach to determine the topological phase boundary. We
derive the general expression for the phase boundary for an arbitrary
probability $P$. Also, our numerical results show that entanglement
entropy is also able to clearly distinguish the topological phase from
the trivial phase. In this paper, we mainly focus on the case of
$P=1/2$. Results for $P=1/3$ are relegated to the
Appendix~\ref{Appendix}.

\section{Analytical study using transfer matrix}
\label{sec:analytical}

We first determine the analytical expressions for the boundary of the
topological–trivial phase transition by focusing on the localization
properties of the zero-energy edge modes. A good quantity to capture the
localization behavior is the Lyapunov exponent $\gamma$, which is
positive for localized states, and vanishes for delocalized states. In
the topological phase, the edge modes are spatially localized on
edges, yielding a positive value for LE; as we approach the phase
boundary, the edge modes merge into the bulk and the LE tends to zero.

To compute $\gamma$ for the edge modes, we employ the transfer-matrix
method, allowing us to clearly identify the phase boundary between the
two insulating phases. Let $\psi_{n,A}$ and $\psi_{n,B}$ be the
wavefunction amplitudes at the $A$ and $B$ sites of the
$n^{\text{th}}$ unit cell, respectively. The Schrödinger equation
$H \psi = E \psi$ gives the coupled equations:
\begin{align}
E \psi_{n,A} &= t_1 \psi_{n,B} + t_2 \psi_{n-1,B}, \label{eq:A_site} \\
E \psi_{n,B} &= t_1 \psi_{n,A} + t_2 \psi_{n+1,A}. \label{eq:B_site}
\end{align}
From Eq.~\eqref{eq:B_site}, we express $\psi_{n+1, A}$ as
\begin{equation}
\psi_{n+1, A} = \frac{E \psi_{n,B} - t_1 \psi_{n, A}}{t_2}.
\label{psi_n_plus_1_A}
\end{equation}
Shifting Eq.~\eqref{eq:A_site} by one lattice site,
\begin{equation}
E \psi_{n+1, A} = t_1 \psi_{n+1, B} + t_2 \psi_{n, B}\nonumber.
\end{equation}
Substituting $\psi_{n+1, A}$ from Eq.~\eqref{psi_n_plus_1_A} into the above gives
\begin{align}
E \left( \frac{E \psi_{n, B} - t_1 \psi_{n, A}}{t_2} \right) &= t_1 \psi_{n+1, B} + t_2 \psi_{n, B}. \nonumber
% t_1 \psi_{n+1, B} &= \frac{E^2 \psi_{n, B} - E t_1 \psi_{n, A}}{t_2} - t_2 \psi_{n, B}\nonumber.
\end{align}
Rewriting, we obtain
\begin{equation}
  \psi_{n+1, B} = -\frac{E}{t_2} \psi_{n, A} + \frac{E^2 - t_2^2}{t_1 t_2} \psi_{n, B}.
  \label{psi_n_plus_1_B}
\end{equation}
Combining Eqs. \eqref{psi_n_plus_1_A} and \eqref{psi_n_plus_1_B}, the dynamics can be expressed in the transfer-matrix form:
\begin{equation}
\begin{pmatrix}
\psi_{n+1, A} \\[0.2cm]
\psi_{n+1, B}
\end{pmatrix}
=T_n
\begin{pmatrix}
\psi_{n, A} \\[0.2cm]
\psi_{n, B}
\end{pmatrix},
\end{equation}
where the transfer matrix is
\begin{equation}
  \label{eq:transfer-matrix}
  T_n = 
  \begin{pmatrix}
    -\dfrac{t_1}{t_2} & \dfrac{E}{t_2} \\[0.5cm]
    -\dfrac{E}{t_2} & \dfrac{E^2 - t_2^2}{t_1 t_2}
  \end{pmatrix}.
\end{equation}
For the states at $E = 0$, the transfer matrix reduces to
\begin{equation}
{T_n}(E=0) = 
\begin{pmatrix}
-\dfrac{t_1}{t_2} & 0 \\[0.3cm]
0 & -\dfrac{t_2}{t_1}
\end{pmatrix}.
\end{equation}

The Lyapunov exponent associated with the zero-energy modes is
defined as
\begin{align}                                                                                                   
\gamma_0 &= \lim_{L \to \infty} \frac{1}{L} \ln \left\|\mathcal{T}_L \right\|,                                                                                                               
\label{Lyapunov_def}                                                                                                                                                                         
\end{align}
where
\begin{equation}
  \label{eq:total-transfer-matrix}
  \mathcal{T}_L = \prod_{n=1}^L T_n
\end{equation}
is the total transfer matrix, and $ \|\mathcal{T}_L\| $ represents
its norm. We take the norm to be the largest absolute eigenvalue of
the transfer matrix.

\subsection{Clean SSH model}
For the clean SSH model, the hopping terms are given by
\begin{align}
  t_1 &= t-\lambda, \nonumber \\
  t_2 &= t+\lambda.
  \label{Clean_Ham}
\end{align}
In this case, since all the transfer matrices
$T_n$ are the same, the total transfer matrix is
\begin{equation}
  \label{eq:total-transfer-matrix-clean}
  \mathcal{T}_L =
  \begin{pmatrix}
    \left( -\dfrac{t_1}{t_2} \right)^L   &  0   \\[0.3cm]
    0  &  \left( -\dfrac{t_2}{t_1} \right)^L
  \end{pmatrix}.
\end{equation}
Here we want to calculate the Lyapunov exponent of the edge
states. For edge states to appear, the system has to be in the
topological regime, i.e. $|t_2|>|t_1|$. Therefore, following
Eqs.~(\ref{Lyapunov_def}) and (\ref{eq:total-transfer-matrix}), the LE
is
\begin{align}
  \gamma_0 = \lim_{L \to \infty} \frac{1}{L} \ln \left|\left( \frac{t_{2}}{t_1} \right)^L \right|
  = \ln \left| \frac{t_2}{t_1} \right|.
\end{align}
In the topological region $\gamma_0>0$ because of the localized nature of the edge states and $\gamma_0 \to 0$ as we approach the phase boundary. Hence the boundary of the topological phase is
\begin{equation}
\left|\frac{t_2}{t_1}\right|=1.
\end{equation}
Since $t_1$ and $t_2$ are given by Eq.~\eqref{Clean_Ham}, the
topological phase boundaries become
\begin{equation}
  \lambda=0~, \quad t=0.
\end{equation}

\begin{figure*}
    \centering
    \includegraphics[width=\textwidth]{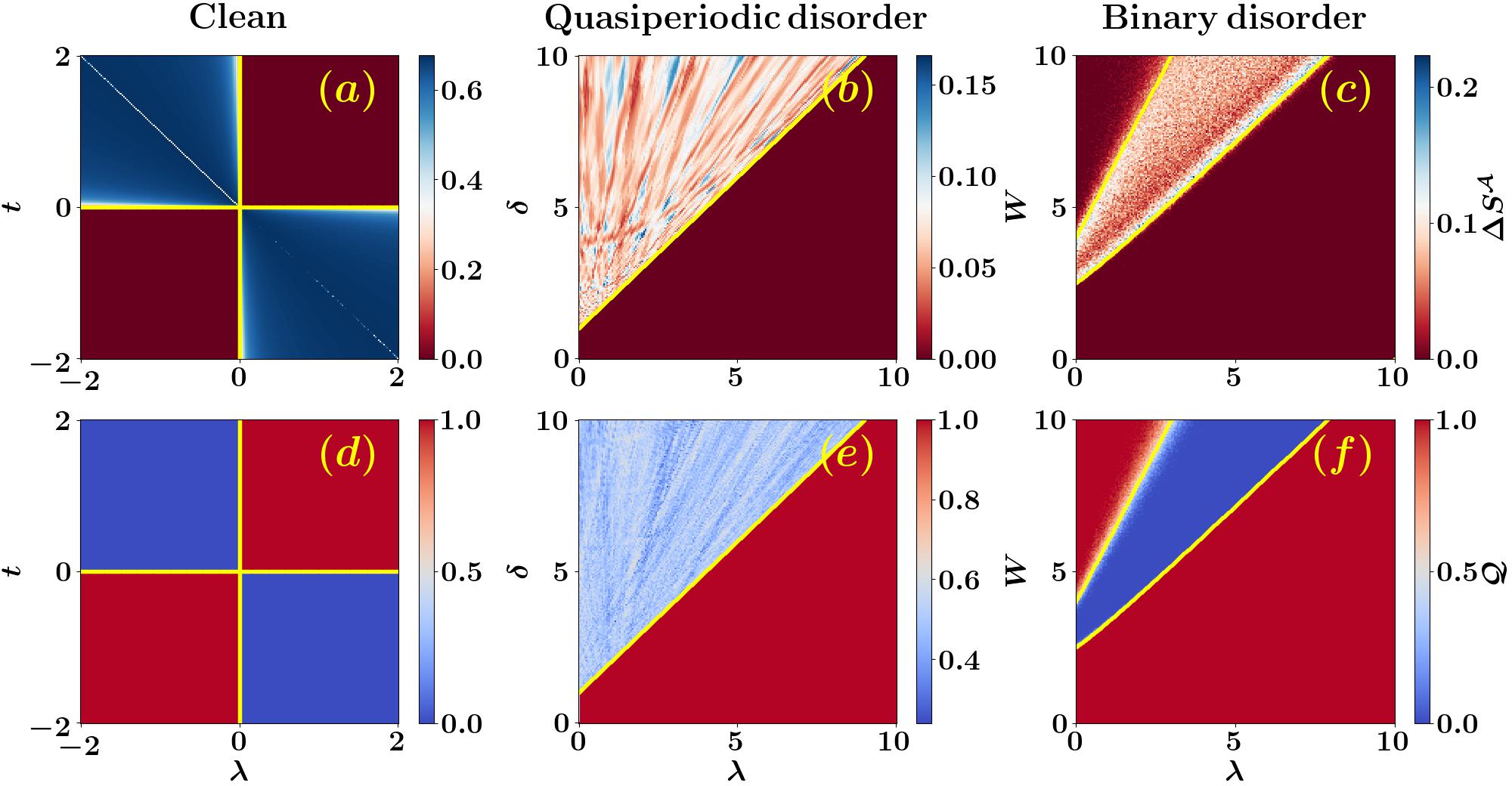}
    
    \caption{Phase diagrams of the SSH model variants. In (a)-(c) the
      color scale denotes
      $\Delta S^{\mathcal{A}} = | S^{\mathcal{A}}_\text{hf} -
      S^{\mathcal{A}}_{\text{hf}+1} |$ with $\text{hf}=N/2$, and the
      subsystem size $N_{\mathcal{A}}=50$. In (d)–(f) the color scale
      denotes topological quantum number $\mathcal{Q}$, which is defined in Appendix \ref{Quantum number}. For the
      disordered cases with $t=1$, the results shown are averaged over
      100 disorder realizations. For all the figures, system size
      $N=400$. Yellow curves mark analytical phase boundaries. On the
      diagonal line $t=-\lambda$ in (a), the intracell hopping vanishes,
      causing the system to break into disconnected
      segments. Therefore the state just above half filling becomes
      localized on a few sites, reducing its contribution to the
      entanglement entropy and hence yielding a smaller
      $ \Delta S^{\mathcal{A}} $.}
    \label{fig:Delta_S_color_plot}
\end{figure*}

\subsection{Quasiperiodic disorder}
In the SSH model with quasiperiodic disorder in the hopping term, the
intercell and the intracell hopping are given by
Eq.~\eqref{quasi_hopping_strength}. Using equation
Eq.~\eqref{eq:total-transfer-matrix}, the total transfer matrix is
\begin{equation}
  \mathcal{T}_L =
  \begin{pmatrix}
    1/Z & 0 \\
    0 & Z
  \end{pmatrix},
  \label{eq:total-transfer-matrix-quasiperiodic}
\end{equation}
where
\begin{equation}
  Z = (-1)^L \prod\limits_{n=1}^L \frac{t+\lambda+\delta \cos(2\pi\beta n+\phi)}{t-\lambda-\delta \cos(2\pi\beta n+\phi)}.
\end{equation}
Here $|Z| > |1/Z|$, the reason for which
will be clear once we obtain the expression for the LE by considering
$|Z|$ as the norm of the transfer matrix.  From
Eqs.~\eqref{Lyapunov_def} and \eqref{eq:total-transfer-matrix} the LE
of the edge states is
\begin{subequations}
  \begin{align}
    \gamma_0  &= \lim_{L \to \infty} \frac{1}{L} \ln \left| Z \right|, \label{eq:gamma-alpha} \\
              &= \lim_{L \to \infty} \frac{1}{L} \ln \left|\prod_{n=1}^L
                \frac{t+\lambda+\delta \cos(2\pi\beta n+\phi)}{t-\lambda-\delta \cos(2\pi\beta n+\phi)} \right|. \label{eq:gamma-quasi}
  \end{align}
\end{subequations}
Since $\beta$ is an irrational number, as $n$ varies, the interval
$[0,2\pi]$ fills uniformly. This follows from Weyl’s equidistribution
theorem and properties of irrational rotations~\cite{Weyl1916,
  irrational}. Then by using the classical Jenson's
formula~\cite{PhysRevB.100.125157, gradshteyn_ryzhik_2007}, we can
write
\begin{align}
  \gamma_0
        &= \frac{1}{2\pi} \int\limits_0^{2\pi}\ln \left |\frac{t+\lambda+\delta \cos\theta}{t-\lambda-\delta \cos\theta}\right | \text{d}\theta \nonumber \\
        &= \frac{1}{2\pi} \int\limits_{0}^{2\pi} \ln|t + \lambda + \delta\cos \theta| \, \text{d}\theta \nonumber \\ 
        & \hspace{6mm} - \frac{1}{2\pi}\int\limits_{0}^{2\pi} \ln|t - \lambda - \delta\cos \theta| \ \text{d}\theta \nonumber .
\end{align}
Both integrals are of the form
\begin{align*}
  \frac{1}{2\pi} \int_{0}^{2\pi} \ln\left|a + b\cos u\right| \, \text{d}u
  &=
    \begin{cases}
      \ln \frac{|a| + \sqrt{a^2 - b^2}}{2}, & \text{if } |a| > |b|, \\
      \ln \frac{|b|}{2}, & \text{if } |a| < |b|.
    \end{cases}
\end{align*}
Using this expression and a careful analysis of the inequalities, the
Lyapunov exponent $\gamma_0$ for the edge state can be written as
\begin{equation}
  \gamma_0 =
  \begin{cases} 
    \ln\frac{t + \lambda + \sqrt{(t + \lambda)^2 - \delta^2}}{|t - \lambda| + \sqrt{(t - \lambda)^2 - \delta^2}},
    & \text{ if } \delta < |t-\lambda|, \\[0.4cm]
    \ln\frac{t + \lambda + \sqrt{(t + \lambda)^2 - \delta^2}}{\delta},
    & \text{ if } |t-\lambda| < \delta < t+\lambda.
\end{cases}
\label{eq:LE_quasiperiodic}
\end{equation}
Clearly, $\gamma_0 > 0$ for $\delta < t + \lambda$, indicating
localized edge states and hence a topological insulating phase. In
contrast, $\gamma_0 \to 0$ as $\delta \to t + \lambda$, therefore the
topological phase boundary is
\begin{equation}
  \delta=t+\lambda.
\end{equation}
This analytical expression is consistent with both our numerical results and earlier numerical studies on the SSH model with quasiperiodic modulation in the hopping amplitudes~\cite{quasi_SSH}.

From Eq.~\eqref{eq:total-transfer-matrix-quasiperiodic} we can see
that the total transfer matrix $\mathcal{T}_L$ has two eigenvalues $Z$
and $1/Z$. The LE obtained in Eq.~\eqref{eq:LE_quasiperiodic} is by
considering the eigenvalue $Z$. Hence, a positive value of $\gamma_0$
indicates that $|Z| > 1$ (according to Eq.~\eqref{eq:gamma-alpha}),
supporting our choice of $Z$ as the largest absolute eigenvalue.

\begin{figure*}
    \centering
    \includegraphics[width=\textwidth]{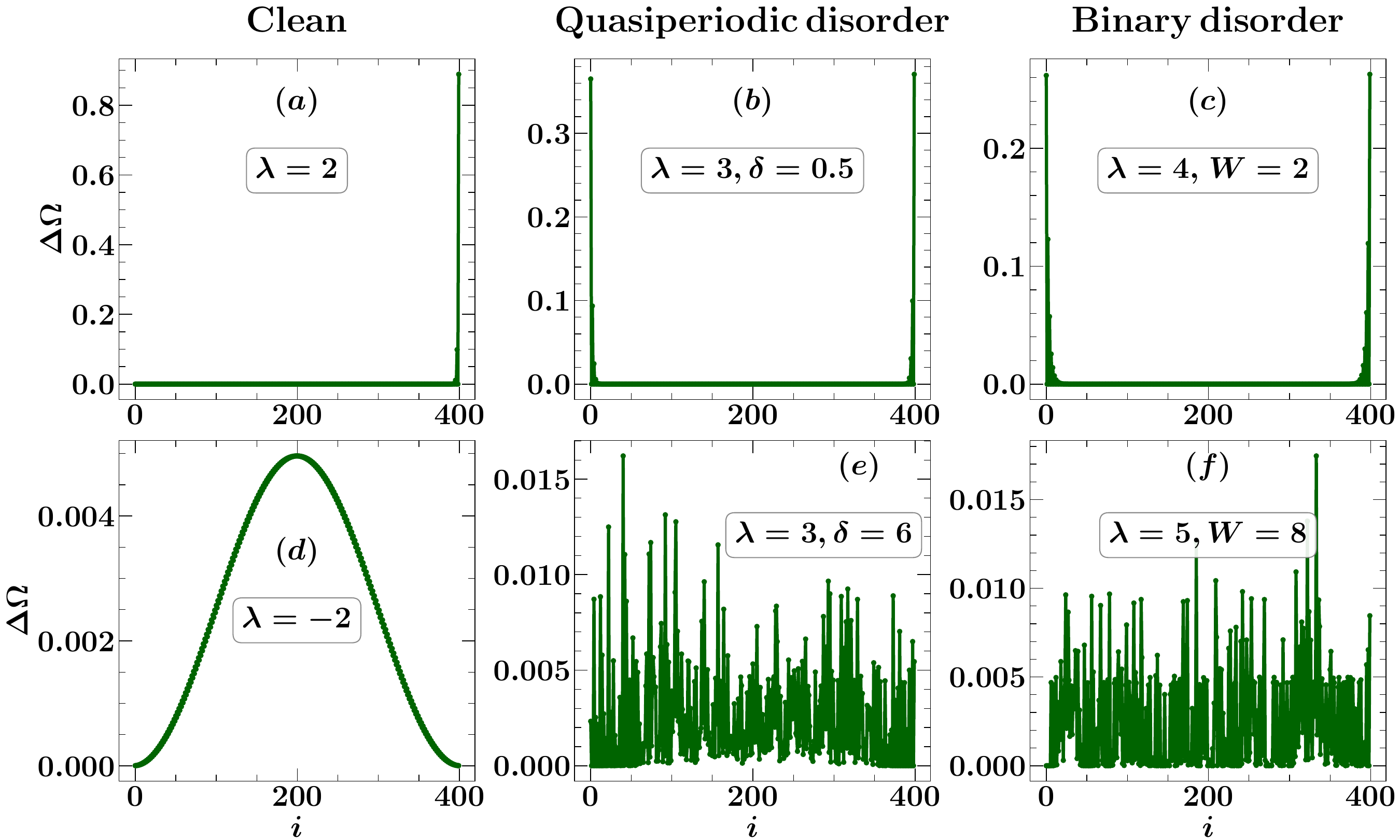}
    \caption{Spatial distribution of
      $\Delta \Omega = \left| \Omega_{\text{hf}} -
        \Omega_{\text{hf}+1} \right|$ for a system of size $N=400$ at
      half-filling ($\text{hf}=200$). The hopping amplitude is $t=1$
      for all cases. Subfigures (a)--(c) correspond to the topological
      phase, while (d)--(f) correspond to the trivial phase.}
    \label{fig:occupation-plot}
\end{figure*}

\subsection{Random binary disorder}
We consider the disordered SSH model where the hopping terms are given
by Eq.~\eqref{random_hopping_strength} and the hopping modulation
$\Delta_n$ is given by Eq.~\eqref{Delta_value}. Here, $\Delta_n$
can take only two values for any $n$, and hence the transfer matrices can take only two possible forms,
\begin{equation}
T_a = \begin{pmatrix}
-\dfrac{v}{u} & 0 \\
0 & -\dfrac{u}{v}
\end{pmatrix}
\end{equation}
corresponding to $\Delta_n = 2$ (with probability $P$), and
\begin{equation}
T_b = \begin{pmatrix}
-\dfrac{v+W}{u-W} & 0 \\
0 & -\dfrac{u-W}{v+W}
\end{pmatrix}
\end{equation}
corresponding to $\Delta_n = 2-W$ (with probability $1-P$), where
$u=t+\lambda+2$ and $v=t-\lambda-2$.  Hence, in
Eq.~\eqref{eq:total-transfer-matrix} for the total transfer matrix,
$T_a$ appears $PL$ number of times, and $T_b$ appears
$(1-P)L$ number of times in the product.  Furthermore, since the
transfer matrices $T_a$ and $T_b$ commute in our case, the total
transfer matrix can be written as
\begin{equation}
\mathcal{T}_L = T_a^{LP} T_b^{L(1-P)}.
\end{equation}
Using Eqs.~(\ref{Lyapunov_def}) and (\ref{eq:total-transfer-matrix})
the LE of the edge states can be written as
\begin{equation}
  \gamma_0= P \ln \|T_a\| + (1-P) \ln \|T_b\|.
\end{equation}
For the edge modes to exist, the intercell hopping has to be larger than
the intracell hopping, i.e., $|t_2|>|t_1|$. Therefore,
$||T_a||=\left|\frac{u}{v}\right|$ and
$||T_b||=\left|\frac{u-W}{v+W}\right|$. Hence,
\begin{align}
\gamma_0 &= P\ln \left| \dfrac{u}{v}\right|
           + (1-P) \ln \left| \dfrac{ u-W }{ v+W }\right| \nonumber \\
&= \ln \left(\left|\dfrac{u}{v} \right|^P  \left|\dfrac{u-W}{v+W} \right|^{1-P}\right) \nonumber \\
&=  \ln \left|\left(\dfrac{u}{v} \right)^P  \left(\dfrac{u-W}{v+W}\right)^{1-P} \right|.
\end{align}
The phase transition occurs when $\gamma_0 = 0$, thus
\begin{equation}
\left|\left(\dfrac{u}{v} \right)^P \cdot \left(\dfrac{u-W}{v+W}\right)^{1-P} \right|=1.
\label{general_P_mid}
\end{equation}
Note that $u>0$ for $t$, $\lambda>0$. However, $v$ can be
negative as well as positive. For $v>0$,
\begin{align}
W_\pm =& \frac{u^{\frac{1}{1-P}} \mp v^{\frac{1}{1-P}}}{u^{\frac{P}{1-P}}\pm v^{\frac{P}{1-P}}},
\label{general_P1}
\end{align}
and for $v<0$,
\begin{align}
W_\pm =& \frac{u^{\frac{1}{1-P}} \mp (-v)^{\frac{1}{1-P}}}{u^{\frac{P}{1-P}}\mp (-v)^{\frac{P}{1-P}}}.
\label{general_P2}
\end{align}
These equations yield the boundaries separating the topological
insulator phase from the trivial insulator phase. We confirm these
boundaries, with a numerical study for $P=1/2$ in the next section
and for $P=1/3$ in the Appendix~\ref{Appendix}.

\section{Entanglement entropy based diagnostic for edge localization}
\label{Numerical}
In this section, we investigate the phase transition between
topological and trivial insulating states using the entanglement
entropy as a diagnostic measure. We demonstrate that the entanglement
entropy is an effective quantity for distinguishing between these two
phases. To gain further insight into the behavior of the entanglement
entropy, we analyze the occupation numbers of the full
system. We contrast our results against those of the
topological invariant $\mathcal{Q}$ in Appendix \ref{Quantum number}. We briefly review these
quantities and present our results.

\subsection{Entanglement Entropy}
\label{entagnlement}

Entanglement entropy (EE) is a quantity used to measure entanglement
in a pure state of a bipartite system. In general, to calculate EE,
one uses the reduced density matrix approach. In this approach, we
first partition the entire system appropriately into subsystems
$\mathcal{A}$ and $\mathcal{B}$. The entanglement entropy of the
subsystem $\mathcal{A}$ can then be written as
$S^{\mathcal{A}}=-\text{Tr}(\rho_{\mathcal{A}} \log
\rho_{\mathcal{A}})$, where
$\rho_{\mathcal{A}}=\text{Tr}_{\mathcal{B}}(\rho)$ is the reduced
density matrix of subsystem $\mathcal{A}$ obtained by tracing out the
degrees of freedom of the subsystem $\mathcal{B}$ from the overall
state $\rho$. In our work, we study the EE of the many-body ground
state $\rho=\ket{\Psi_0}\bra{\Psi_0}$. To obtain this using exact
diagonalization, we first have to diagonalize the Hamiltonian (which
is a $2^{N} \times 2^{N}$ matrix, with $N=2L$ being the total number
of sites) to find the ground state $\ket{\Psi_0}$, and then computing
$S^{\mathcal{A}}$ involves diagonalizing $\rho_{\mathcal{A}}$.

For free fermionic systems, Peschel et. al.~\cite{Ingo_Peschel_2003,
  Peschel_2009, Peschel2012} developed an approach, the correlation
matrix approach, that drastically reduces the computation time. The
details of the method are as follows.  For non-interacting fermionic
systems, since the ground state has the structure of a Slater
determinant, Wick's theorem can be exploited to express the reduced
density matrix in the form
$\rho_{\mathcal{A}}=\text{e}^{-H_{\mathcal{A}}}/Z$, where
$H_{\mathcal{A}}$ is the entanglement Hamiltonian and $Z$ is a
normalization factor ensuring $\text{Tr}(\rho_{\mathcal{A}})=1$.  The
correlation matrix $C$ for the full system is defined as
$C_{ij} = \langle c_i^\dagger c_j \rangle$ and has dimension
$N \times N$. Subsequently, one selectively extracts the portion of
the correlation matrix that pertains to the subsystem of interest and
the correlation matrix of the subsystem $\mathcal{A}$,
i.e. $C^{\mathcal{A}}$ becomes of the order of $L\times L$. The
correlation matrix, $C^{\mathcal{A}}$ and the entanglement Hamiltonian
$H_{\mathcal{A}}$ are related as~\cite{Ingo_Peschel_2003,
  Peschel_2009, Peschel2012}
\begin{equation}
  C^{\mathcal{A}}=\frac{1}{e^{H_{\mathcal{A}}}+1}.
\end{equation}
The entanglement entropy between the subsystems $\mathcal{A}$ and
$\mathcal{B}$ for free fermions is determined using eigenvalues
$\lambda_k$ of the subsystem correlation matrix by~\cite{Peschel_2009,
  Peschel2012}
\begin{equation}
   S^{\mathcal{A}}=-\sum_k[\lambda_k\ln \lambda_k+(1-\lambda_k)\ln (1-\lambda_k)].
   \label{eq:entropy}
\end{equation}
This method provides an efficient way to compute the entanglement
entropy of the many particle ground state of free fermions and has
been extensively used in recent works for various models.

We demonstrate that the EE of the many-body ground state is an
effective probe of the quantum phase transition between the
topological and trivial (band) insulating phases. We compute the
difference in EE between the many-particle ground state at
half-filling and the state with one more than half filling, defined as
\begin{equation}
\Delta S^{\mathcal{A}} = \left| S^{\mathcal{A}}_\text{hf} - S^{\mathcal{A}}_\text{hf+1} \right|.
\end{equation}
To utilize EE as a diagnostic measure, it is crucial to define the
subsystem appropriately. We find that choosing a subsystem
$\mathcal{A}$ consisting of a few unit cells deep within the bulk as
shown in the schematic (Fig.~\ref{fig:subsystem_A}), and the remaining
sites as subsystem $\mathcal{B}$, is an effective choice. The
rationale for this choice will become clear in the subsequent
analysis. We note that considering the difference with respect to a
state with one particle fewer than half-filling (
$\left|S^{\mathcal{A}}_\text{hf} -
  S^{\mathcal{A}}_\text{hf-1}\right|$) also yields the same
qualitative behavior across the phase transition. Here we focus on the
results for one particle added beyond half-filling.

%%%%%%%%%%%%%%%%%%%%%%%%%%%%%%%%%%%%%%%%%%%%%%%%%%%%%%%%%%%%%%%%%%%%%%%%%%%%%%%%%%%%%%%%%%%%%%%%%% Clean SSH Model %%%%%%%%%%%%%%%%%%%%%%%%%%%%%%%%%%%%%%%%%%%%%%%%%%%%%%%%%%%%%%%%%%%%%%%%%%%%%%%%%%%%%%%%%%%%%%%%%%%%%%%%%%%

As discussed earlier, the clean SSH model (Eq.~\eqref{Clean_Ham}),
exhibits a topological phase transition from a trivial to a
topological insulator, depending on the relative strengths of
intercell and intracell hopping amplitudes. We plot the phase diagram
of $t$ and $\lambda$ with $\Delta S^{\mathcal{A}}$ in
Fig.~\ref{fig:Delta_S_color_plot}\flc{a}. In the topological region,
we find $\Delta S^{\mathcal{A}} = 0$, implying that the EE of the
half-filled ground state is identical to that of the ground state with
one additional particle. In contrast, within the trivial phase,
$\Delta S^{\mathcal{A}} \ne 0$, providing a clear distinction from the
topological phase. We further test our quantity
$\Delta S^{\mathcal{A}}$ for the SSH model with the quasiperiodic
disorder in the hopping strength
(Eq.~\eqref{quasi_hopping_strength}). The boundary separating the
topological phase from the trivial phase is $\delta=t+\lambda$. The
quantity $\Delta S^{\mathcal{A}}$ can clearly distinguish two phases
even for the quasiperiodic model as can be seen in
Fig.~\ref{fig:Delta_S_color_plot}\flc{b}. To further confirm that
$\Delta S^{\mathcal{A}}$ effectively identifies the phase transition,
we check this on the SSH model with random binary disorder in hopping
strength (Eq.~\eqref{Delta_value}) for $P=1/2$. The expression for the
phase transition boundary for this case can be obtained by putting
$P=1/2$ in Eq.~\eqref{general_P1} and Eq.~\eqref{general_P2}. We plot
the phase diagram for $t=1$ and our numerical results are in exact
agreement with the analytical expression as shown in
Fig. ~\ref{fig:Delta_S_color_plot}\flc{c}. This confirms that our
quantity $\Delta S^{\mathcal{A}}$ successfully identifies both the
boundaries of the topological regions, in agreement with the
analytical results. It is zero in the topological phase and nonzero in
the trivial phase. We also show that this method is robust by
considering other parameter values in the Appendix~\ref{Appendix}. Also in section~\ref{Casual_localized states}, we discuss how our diagnostic
fares for topologically trivial states localized at the edges.

\begin{figure}
    \centering
    \includegraphics[width=0.48\textwidth]{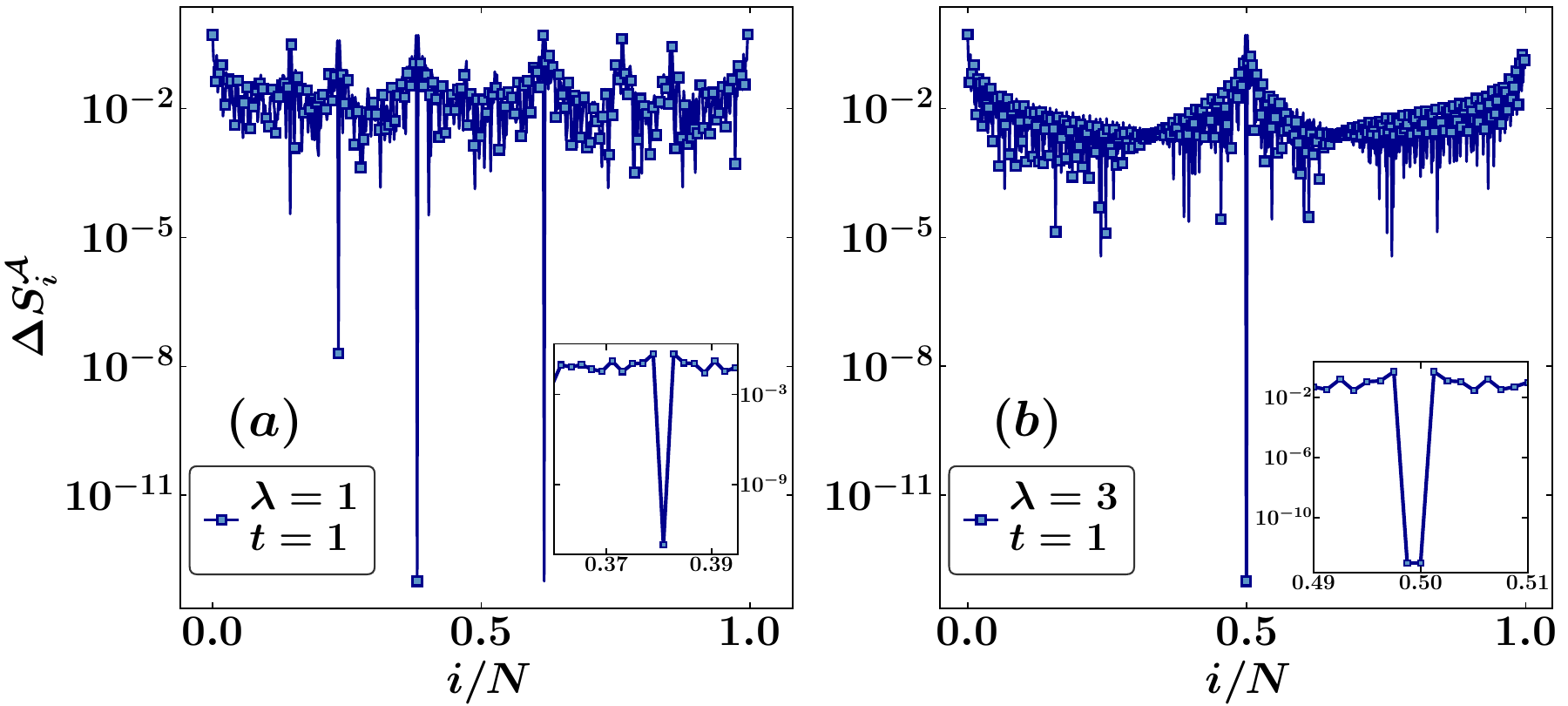}
    \caption{Entanglement-entropy difference
      $\Delta S^{\mathcal{A}}_i = \left| S^{\mathcal{A}}_i -
        S^{\mathcal{A}}_{i+1} \right|$ for all eigenstates in (a) the
      AAH model and (b) the clean SSH model. In (a), the inset shows
      that only a single eigenstate exhibits vanishing
      $\Delta S^{\mathcal{A}}_i$, whereas in (b), the SSH model
      features pairs of states with $\Delta S^{\mathcal{A}}_i = 0$,
      consistent with its topological edge modes.}
    \label{DeltaS_vs_index}
\end{figure}

\subsection{Occupation number}
\label{Occupation}
In the topological regime, $\Delta S^{\mathcal{A}} = 0$ indicates that
the additional particle introduced beyond half-filling does not
contribute to the entanglement entropy of the subsystem
$\mathcal{A}$. To investigate this further, we examine the spatial distribution of the added particle by analyzing the site occupations of the full system. These site occupations correspond to the diagonal elements of the full system's correlation matrix, $C$. In particular,
we consider the change in site occupations defined as
\begin{equation}
\Delta \Omega = \big| \Omega_\text{hf} - \Omega_\text{hf+1} \big|,
 \label{Delta_Omega}
\end{equation}
where $\Omega$ denotes the set of site occupations, corresponding to
the diagonal elements of the full system's correlation matrix. In this
way, we obtain direct information about the location of the additional
particle introduced beyond half-filling.

We study the spatial distribution of $\Delta \Omega$ across the full
system for the clean SSH model in
Fig.~\ref{fig:occupation-plot}\flc{a, d}, the quasiperiodic disorder
case in Fig.~\ref{fig:occupation-plot}\flc{b, e}, and the random
binary disorder case in Fig.~\ref{fig:occupation-plot}\flc{c, f}. From
Figs.~\ref{fig:occupation-plot}\flc{a-c} it can be observed that in
the topological phase, $\Delta \Omega$ is nonzero only at the edges,
indicating that the additional particle introduced beyond half-filling
is localized exclusively at the boundaries. As a result, it does not
occupy sites within subsystem $\mathcal{A}$ and there is no change in
the eigenvalues of the correlation matrix of subsystem $\mathcal{A}$,
consequently, does not contribute to the entanglement entropy. In
contrast, in the trivial phase, the added particle occupies the sites
in both the subsystems $\mathcal{A}$ and $\mathcal{B}$, as seen in
Figs.~\ref{fig:occupation-plot}\flc{d-f}, leading to a change in the
eigenvalues of $C_\mathcal{A}$ and hence finite contribution to the
entanglement entropy, consequently leading to a nonzero value of
$\Delta S^{\mathcal{A}}$. In all the three cases also, the additional
particle is localized at the edges in the topological regime, while in
the trivial regime it occupies both the edges and the bulk, yielding
consistent behavior irrespective of the presence and nature of the
disorder.

\subsection{Non-topological localized edge states away from half-filling}
\label{Casual_localized states}
We discuss the possibility of observing states localized at the edges
that are not of topological origin. For this purpose, we consider the
Aubry--Andr\'e--Harper (AAH) model, a topologically trivial system
that undergoes a transition from a fully extended phase to a fully
localized phase as the strength of the quasiperiodic potential is
increased~\cite{aubry1980}. The model Hamiltonian is
\begin{align*}
H = -J\sum_{n=1}^{L} \left( c_n^\dagger c_{n+1} + \text{H.c.} \right)
+ \sum_{n=1}^{L} V_n\, c_n^\dagger c_n ,
\end{align*}
where $ V_n = \lambda \cos\left[2\pi(\beta n + \theta)\right] $ and
$ \lambda $ is the potential strength. In this model, all
eigenstates are delocalized for $ \lambda < 2t $, whereas they
become localized for $ \lambda > 2t $. It has been observed that
even in the delocalized regime, one occasionally finds eigenstates
that appear localized~\cite{PhysRevB.100.195143}. Looking at their
wavefunction profile we observe that these states are localized at the
edges despite the system being topologically trivial.

We analyse these states using our entanglement entropy based quantity
for the entire spectrum. We compute
\[
\Delta S^{\mathcal{A}}_i = \left| S^{\mathcal{A}}_i -
S^{\mathcal{A}}_{i+1} \right|,
\]
where $ i $ labels the filling across the entire spectrum and
$\mathcal{A}$ is the subsystem consisting of a few unit cells deep inside
the bulk. For the AAH model, we find that
$ \Delta S^{\mathcal{A}}_i $ indeed vanishes for those eigenstates
that are localized at the edges
[Fig.~\ref{DeltaS_vs_index}\flc{a}]. However, these states are not
topological as for a state to be topologically protected, it must
appear at half-filling at zero energy, which is the requirement of
chiral symmetry. In contrast, in the SSH model, not only states with
vanishing $\Delta S^{\mathcal{A}}$ come at half filling but also come
in pairs i.e., $ \Delta S^{\mathcal{A}}_i $ as well as
$ \Delta S^{\mathcal{A}}_{i+1} $ both are zero
[Fig.~\ref{DeltaS_vs_index}\flc{b}], consistent with their genuine
topological origin. For AAH model, states with vanishing
$\Delta S^{\mathcal{A}}$ do not come in pairs, i.e.
$ \Delta S^{\mathcal{A}}_{i+1}\neq 0 $. This discussion shows that 
the diagnostic we propose primarily identifies edge-localized states. 
To use it to characterize a topological phase transition, one also 
has to consider the physical context.

\begin{figure}
    \centering
    \includegraphics[width=0.48\textwidth]{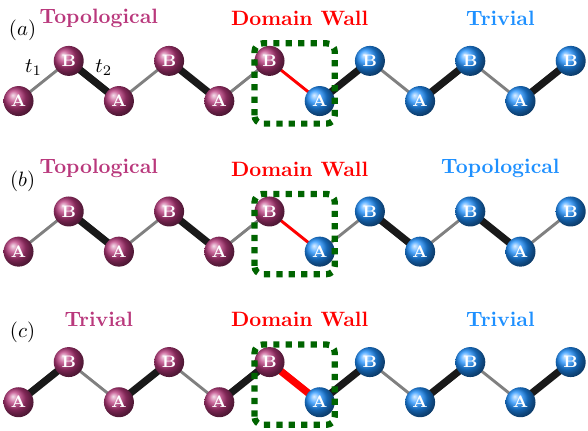}
    \caption{Schematic illustration of situations in which domain
      walls arise in the SSH chain: (a) a topological SSH chain joined
      to a trivial SSH chain by a weak bond; (b) two topological SSH
      chains joined by a weak bond; and (c) two topologically
      trivial SSH chains joined by a strong bond.}
\label{Domain_wall_schematic}
\end{figure}

\begin{figure*}
    \centering
    \includegraphics[width=\textwidth]{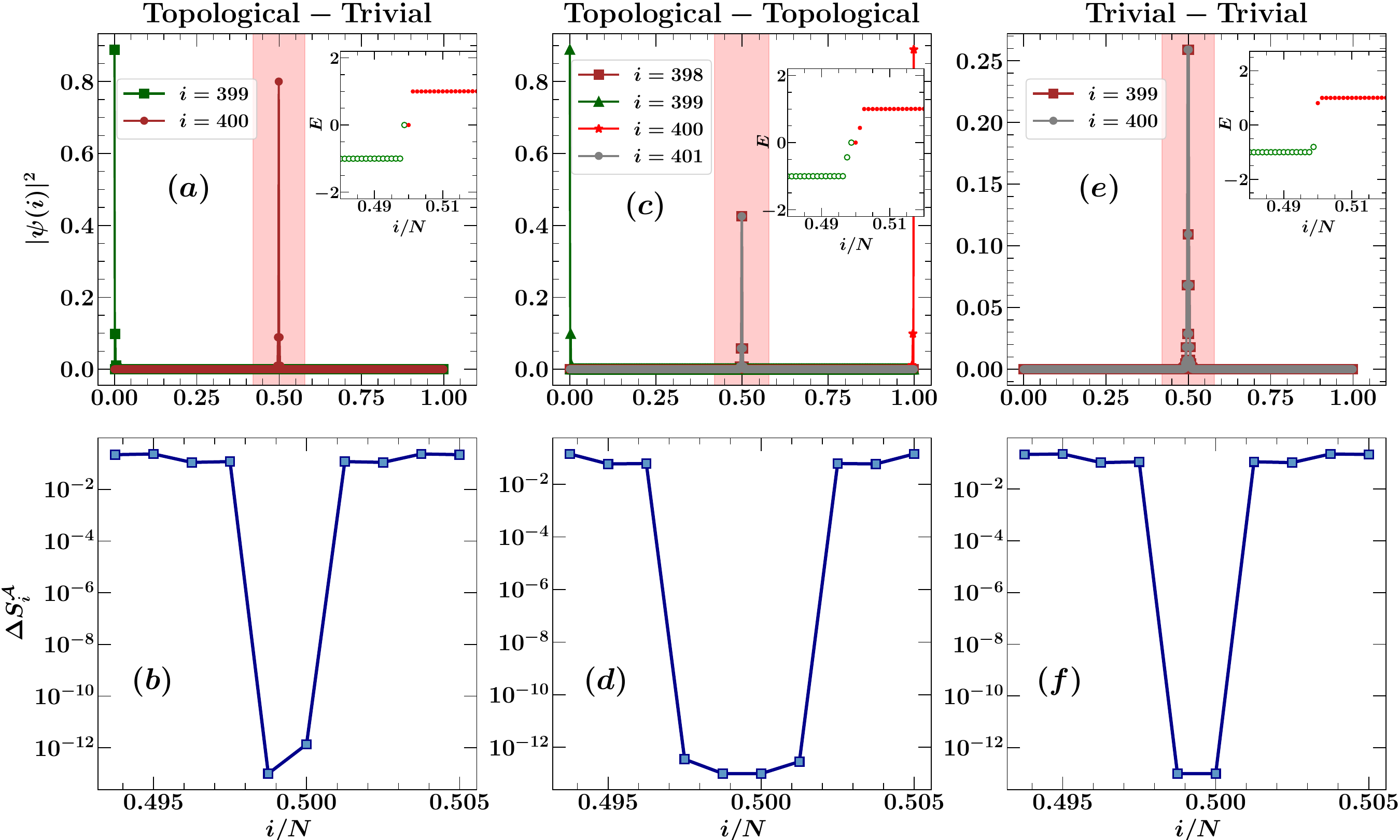}
    \caption{Trivial-topological, topological-topological, and
      trivial-trivial junctions in the SSH chain.  (a,c,e) Probability
      density $|\psi(i)|^2$ of localized states for (a) a
      trivial-topological junction, (c) a topological-topological
      junction, and (e) a trivial-trivial junction.  In panel (a), one
      zero-energy state is localized at the left edge and another at
      the domain wall.  In panel (c), four localized states appear:
      two zero-energy edge states and two finite-energy states
      localized at the domain wall.  In panel (e), two finite-energy
      states are localized at the domain wall. The red shaded region
      denotes subsystem $\mathcal A$ with $N_{\mathcal A}=128$, and
      the insets show the central part of the energy spectra.  (b,d,f)
      Corresponding entanglement entropy difference
      $\Delta S^{\mathcal A}_i = |S^{\mathcal A}_i - S^{\mathcal
        A}_{i+1}|$ as a function of filling near half filling for the
      respective junctions. The vanishing values of
      $\Delta S^{\mathcal A}_i$ correspond to the localized states
      shown in the upper panels. For topological SSH chain $t_1=0.5$
      and $t_2=1.5$, and for the trivial SSH chain $t_1=1.5$ and
      $t_2=0.5$. The system size is $N=800$ in all the figures.}

    \label{Domain_wall}
\end{figure*}

\begin{figure*}
    \centering
    \includegraphics[width=\textwidth]{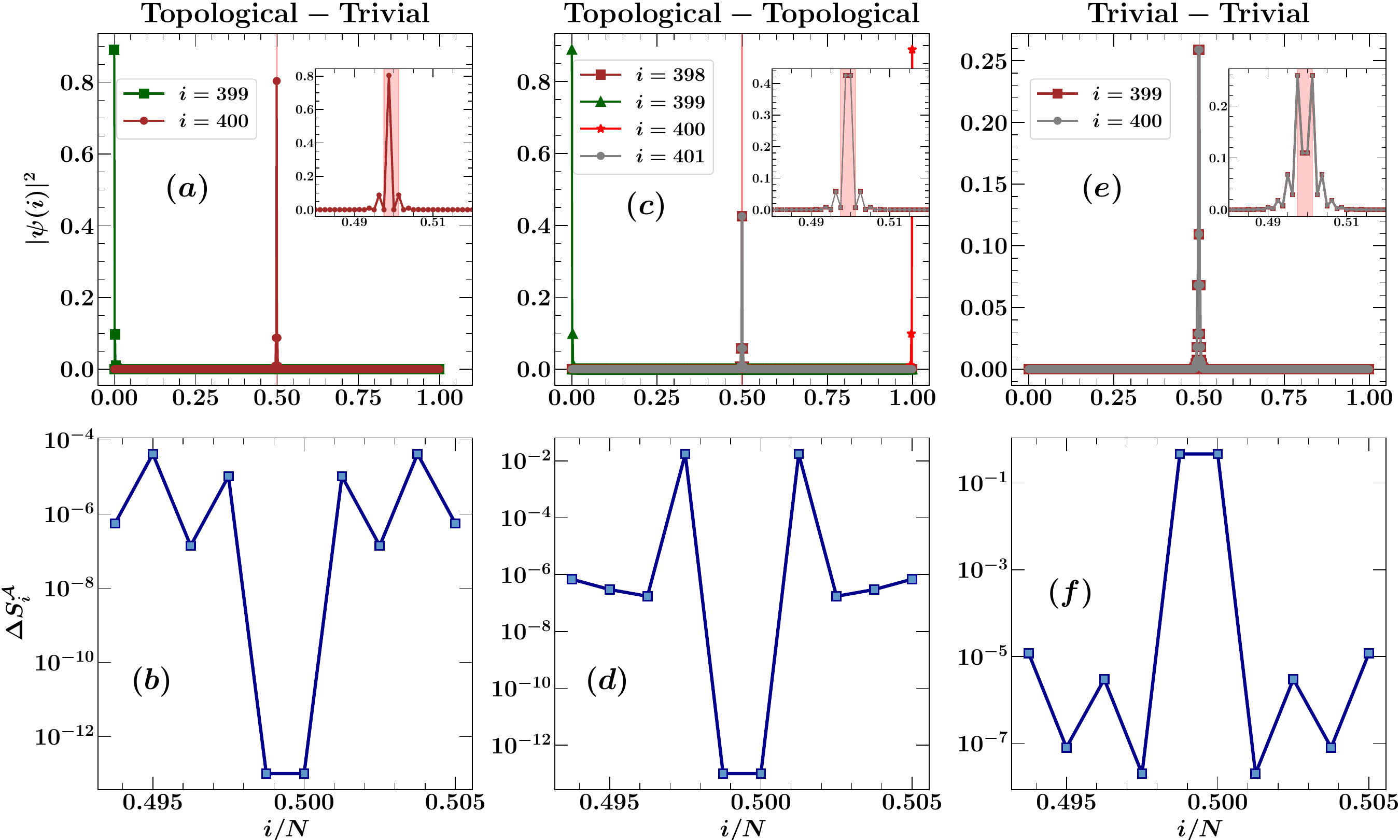}
\caption{Same as Fig.~\ref{Domain_wall}, but with subsystem 
$\mathcal A$ chosen as $N_{\mathcal A}=4$. 
The red shaded region denotes the modified subsystem, and the insets 
show a zoomed-in view highlighting the subsystem choice. 
All other parameters are identical to those in Fig.~\ref{Domain_wall}.}

    \label{Domain_wall2}
\end{figure*}

There exist accidental situations where states localized at the edges
of a system appear in pairs at $E=0$ and at half filling without
having a topological origin, for example in superconducting wires
hosting Majorana modes. Although we do not encounter such states in
our present models, they can arise in the superconducting
wires. However, these edge-localized states are not robust against
changes in system parameters: modifying the parameters typically leads
these edge states to merge with the bulk. This behavior can be
exploited by altering the length of the topological region in a
superconducting wire. If the edge states have a genuine topological
origin, and the few sites are tuned to be away from the topological
origin, then the edge states should still survive at the new
topological boundary with $E=0$, and hence $\Delta S^{\mathcal{A}}$
continues to vanish for them. On the other hand, spurious zero-energy
edge states split and merge into the bulk under such modifications,
resulting in a nonzero $\Delta S^{\mathcal{A}}$. This protocol has
been discussed in detail in the context of Majorana edge
modes~\cite{PhysRevB.108.205426}, although no entanglement based
measures are studied here. 

There exists another class of configurations in which non-topological
domain-wall states emerge near half filling. In the following section,
we analyze this case in detail and outline a systematic approach to
distinguish these trivial states from genuinely topological
domain-wall states.

\begin{figure}
    \centering
    \includegraphics[width=0.48\textwidth]{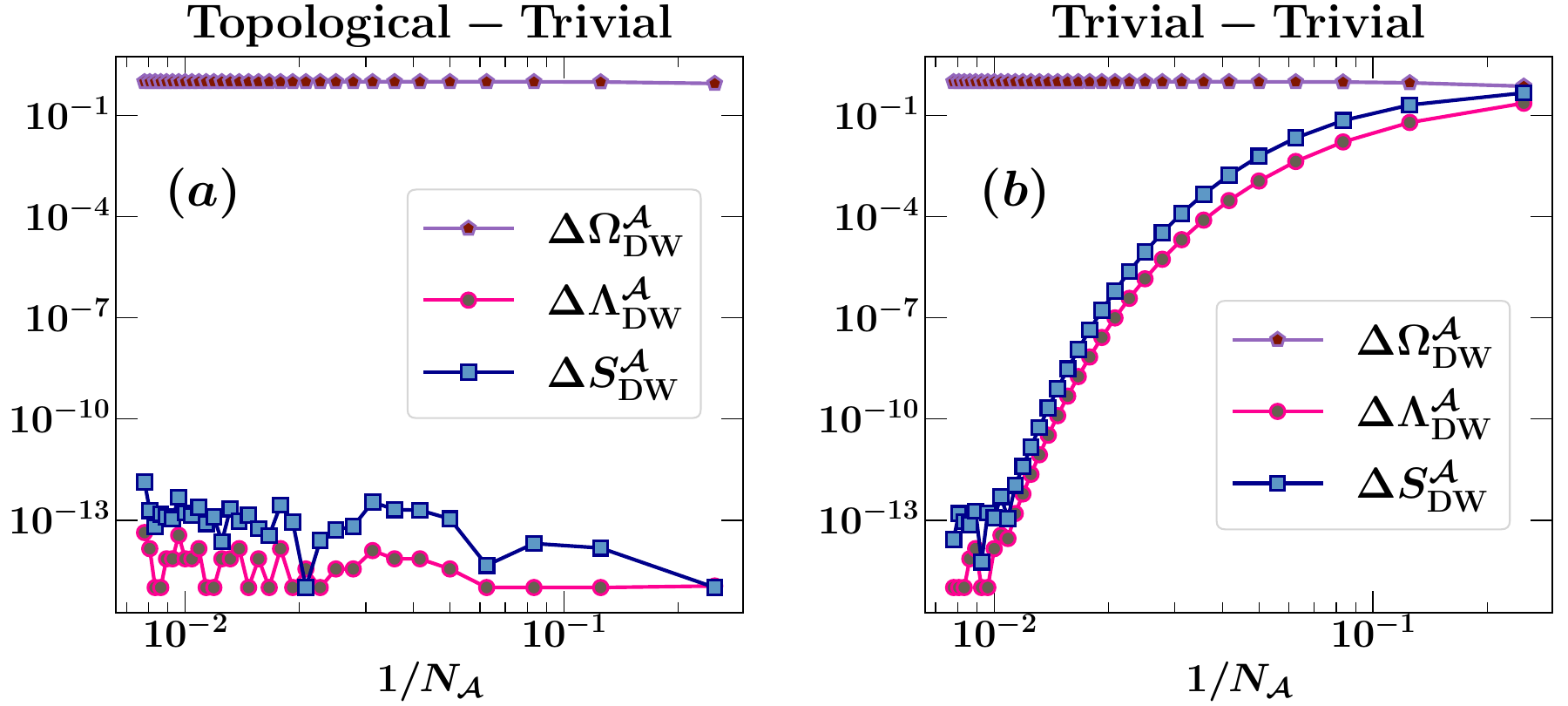}
    \caption{Behavior of different quantities as the subsystem size
      $N_{\mathcal A}$ is varied for domain-wall states. The
      horizontal axis shows $1/N_{\mathcal A}$, so moving to the right
      corresponds to shrinking the subsystem. Figure (a) shows a
      topological domain wall in a topological--trivial junction,
      while figure (b) shows a trivial domain wall in a
      trivial--trivial junction. The plotted quantities are the change
      in entanglement entropy $\Delta S^{\mathcal A}_{\mathrm{DW}}$,
      the change in the spectral measure
      $\Delta \Lambda^{\mathcal A}_{\mathrm{DW}}$, and the change in
      subsystem occupation $\Delta \Omega^{\mathcal A}_{\mathrm{DW}}$
      when a particle is added to the domain-wall state. In the
      topological case (a), $\Delta S^{\mathcal A}_{\mathrm{DW}}$
      remains essentially zero as the subsystem is reduced, indicating
      robustness of the topological state. In contrast, for the
      trivial case (b), $\Delta S^{\mathcal A}_{\mathrm{DW}}$
      increases and becomes finite as the subsystem shrinks. The
      occupation change $\Delta \Omega^{\mathcal A}_{\mathrm{DW}}$
      remains nearly constant in both cases. For the topological SSH
      chain $t_1=0.5$, $t_2=1.5$; for the trivial SSH chain $t_1=1.5$,
      $t_2=0.5$. The system size is $N=800$, and the initial subsystem
      size is $N_{\mathcal A}=128$.  }
    \label{with_NA}
\end{figure}

\section{Domain-Wall States: Distinguishing Topological from Trivial Localization}
\label{Domain_wall_section}

Up to this point, the diagnostic we propose has relied on the
localization properties of edge states: by detecting states localized
near the system boundaries close to half filling, we identify regimes
with topological character. However, there exist situations in which
localized states appear near half filling without having a topological
origin. The domain-wall states of the SSH chain are a striking example
of this kind. In these situations, an additional level of analysis is
required to correctly identify the phase.

\subsection{Domain-Wall States in the SSH Chain}

Depending on how two SSH chains are joined, different types of domain
walls may arise, as illustrated schematically in
Fig.~\ref{Domain_wall_schematic}. When a topological SSH chain is
connected to a trivial SSH chain by a weak bond
[Fig.~\ref{Domain_wall_schematic}\flc{a}], two zero-energy states
appear: one localized at the left edge of the system and the other
localized at the domain-wall position
[Fig.~\ref{Domain_wall}\flc{a}]~\cite{Munoz2018}. When using
$\Delta S^{\mathcal A}$ as a diagnostic, we choose the subsystem
$\mathcal A$ according to our prescription, such that the additional
particle occupies only one of the two subsystems. This can be achieved
by selecting $\mathcal A$ so that the domain wall lies entirely either
inside or outside the chosen subsystem. With this choice, the
vanishing contributions to $\Delta S^{\mathcal A}_i$, when evaluated
near half filling, occur in pairs and correspond to the genuine
topological edge states, as shown in Fig.~\ref{Domain_wall}\flc{b}.

We next analyze junctions formed by topological--topological and
trivial--trivial SSH chains. When two topological SSH chains are
connected by a weak bond [Fig.~\ref{Domain_wall_schematic}\flc{b}], in
addition to the two zero-energy topological edge states, two
finite-energy states localized at the domain wall appear
[Fig.~\ref{Domain_wall}\flc{c}], which do not have a topological
origin. The corresponding entanglement entropy difference
$\Delta S^{\mathcal A}_i$, evaluated near half filling, exhibits four
dips [Fig.~\ref{Domain_wall}\flc{d}], reflecting the localized nature
of these states and potentially indicating a topological signal when
the subsystem is chosen according to the above
prescription. Similarly, when two trivial SSH chains are connected by
a strong bond [Fig.~\ref{Domain_wall_schematic}\flc{c}], two
finite-energy, topologically trivial states localized at the domain
wall appear [Fig.~\ref{Domain_wall}\flc{e}]. In this case as well,
$\Delta S^{\mathcal A}_i$ displays two dips near half filling
[Fig.~\ref{Domain_wall}\flc{f}], since the added particle
predominantly occupies a single subsystem. In both cases,
$\Delta S^{\mathcal A}_i = 0$ for domain-wall states without
topological origin, for the same reason discussed earlier: the added
particle resides within a single chosen subsystem.

\subsection{Subsystem Tuning as a Topological Discriminator}
In such scenarios, confirming the topological nature of the phase
requires an additional step beyond the mere detection of edge
localization near half-filling. This is achieved by tuning the chosen
subsystem. The red shaded region in Fig.~\ref{Domain_wall} indicates
the initially selected subsystem \( \mathcal{A} \). To test the
robustness of the diagnostic, we shrink the subsystem
\( \mathcal{A} \) such that the particle associated with the
domain-wall state has nonzero occupation in both subsystems, as
illustrated in the insets of Figs.~\ref{Domain_wall2}\flc{a},
\ref{Domain_wall2}\flc{c}, and \ref{Domain_wall2}\flc{e}. For genuine
topological zero-energy states, \( \Delta S^{\mathcal A}_i \) remains
vanishing even under this modified choice of subsystem
[Fig.~\ref{Domain_wall2}\flc{b}].  In contrast, for this choice of
subsystem, no dip appears in \( \Delta S^{\mathcal A}_i \) for
non-topological finite-energy domain-wall states
[Figs.~\ref{Domain_wall2}\flc{d} and \ref{Domain_wall2}\flc{f}].  A
comparison between Figs.~\ref{Domain_wall} and \ref{Domain_wall2}
further shows that partially enveloping the domain wall reduces the
number of dips in \( \Delta S^{\mathcal A}_i \) only for
non-topological domain-wall states. This demonstrates that the
proposed diagnostic can distinguish zero-energy topological states
from finite-energy trivial ones, even when both are spatially
localized. To further understand this behavior, we examine how
$ \Delta S^{\mathcal A}_{\mathrm{DW}}= |S^{\mathcal A}_{\text{DW}} -
S^{\mathcal A}_{\text{DW}-1}| $, corresponding to the case where the
added particle occupies the domain-wall state, evolves as the
subsystem size is reduced. As shown in Fig.~\ref{with_NA}, for a
topological-trivial junction, where the domain-wall state is
topological and pinned at \( E = 0 \),
\( \Delta S^{\mathcal A}_{\mathrm{DW}} \) remains zero upon decreasing
the subsystem size. In contrast, for a trivial-trivial junction, where
the domain-wall state has finite energy,
\( \Delta S^{\mathcal A}_{\mathrm{DW}} \) increases and becomes finite
as the subsystem is shrunk.

To understand the origin of this behavior, we analyze the changes in
the eigenvalues of the subsystem correlation matrix
\( C_{\mathcal A} \). From Eq.~\eqref{eq:entropy}, the contribution to
the entanglement entropy from each eigenvalue vanishes when
\( \lambda_k = 0 \) or \( 1 \), and is maximal at
\( \lambda_k = 0.5 \). The change in entanglement entropy upon adding
a particle can therefore be understood by tracking the redistribution
of eigenvalues relative to \( 0.5 \). We define
\begin{equation}
\Lambda^\mathcal{A}_i = \sum_k \left| \lambda_k - \frac{1}{2} \right|,
\end{equation}
which measures the distribution of $\lambda_k$s relative to 0.5, where
\( i \) denotes the filling. We further define
\( \Delta \Lambda^\mathcal{A}_i = |\Lambda^\mathcal{A}_{\text{DW}} -
\Lambda^\mathcal{A}_{\text{DW}+1}| \).  A vanishing
\( \Delta \Lambda^\mathcal{A} \) may arise either when the eigenvalues
remain unchanged or when they shift symmetrically about \( 0.5 \). To
contrast this spectral measure with an occupation-based quantity, we
also consider $\Omega^{\mathcal{A}}_i = \sum_k \lambda_k$, which
represents the total occupation of subsystem $\mathcal{A}$.  We
further define
$\Delta \Omega^{\mathcal{A}}_{\text{DW}} =
|\Omega^\mathcal{A}_{\text{DW}} - \Omega^\mathcal{A}_{\text{DW}+1}|$,
which quantifies the fraction of the added particle entering subsystem
$\mathcal{A}$.

We compute $\Delta \Omega^\mathcal{A}_{\mathrm{DW}}$ and
$\Delta \Lambda^\mathcal{A}_{\mathrm{DW}}$ as functions of subsystem
size in Fig.~\ref{with_NA} and compare them with
$\Delta S^\mathcal{A}_{\mathrm{DW}}$ as $N_{\mathcal A}$ is reduced.
In both Figs.~\ref{with_NA}\flc{a} and \ref{with_NA}\flc{b},
$\Delta \Omega^\mathcal{A}_{\mathrm{DW}}$ remains nearly unchanged
upon decreasing $N_{\mathcal A}$, indicating that the fraction of the
added particle residing in subsystem $\mathcal A$ is similar for both
topological and trivial domain-wall states. This shows that the
occupation fraction alone cannot account for the distinct behavior of
$\Delta S^\mathcal{A}_{\mathrm{DW}}$. In contrast,
$\Delta \Lambda^\mathcal{A}_{\mathrm{DW}}$ remains zero
[Fig.~\ref{with_NA}\flc{a}] for the topological domain wall as
$N_{\mathcal A}$ decreases, whereas it becomes finite for the trivial
case [Fig.~\ref{with_NA}\flc{b}]. In fact the graph of
$\Delta\Lambda^\mathcal{A}_{\mathrm{DW}}$ closely mimics that of
$\Delta S^\mathcal{A}_{\mathrm{DW}}$ as shown in
Fig.~\ref{with_NA}\flc{b}. We note that
$\Delta S^\mathcal{A}_{\mathrm{DW}}$ remains zero as the subsystem
size is shrunk although $\Delta \Omega^\mathcal{A}_{\mathrm{DW}}$
remains unchanged. This indicates that
$\Delta S^\mathcal{A}_{\mathrm{DW}}$ remaining zero as the subsystem
size is shrunk is not just a consequence of the topological state
having very tiny localization length. In contrast to the cases
described in the previous section, the added particle induces
non-trivial changes in the eigenvalues of the subsystem correlation
matrix.  However, in the topological configuration, the redistribution
of eigenvalues is symmetric around $0.5$, leading to
$\Delta S^{\mathcal A}_{\mathrm{DW}} \approx 0$. Thus we observe that
a characteristic feature of topological domain wall states is that
$\Delta S^{\mathcal A}_{\mathrm{DW}}$ remains zero as the subsystem is
shrunk.

% The characteristic feature of a
% topologically protected state is its robustness against
% perturbations.

% We observe here that this robustness carries over to
% $\Delta S^{\mathcal A}$, which remains zero as the subsystem size is
% reduced.

  \begin{figure*}
    \centering
    \includegraphics[width=\textwidth]{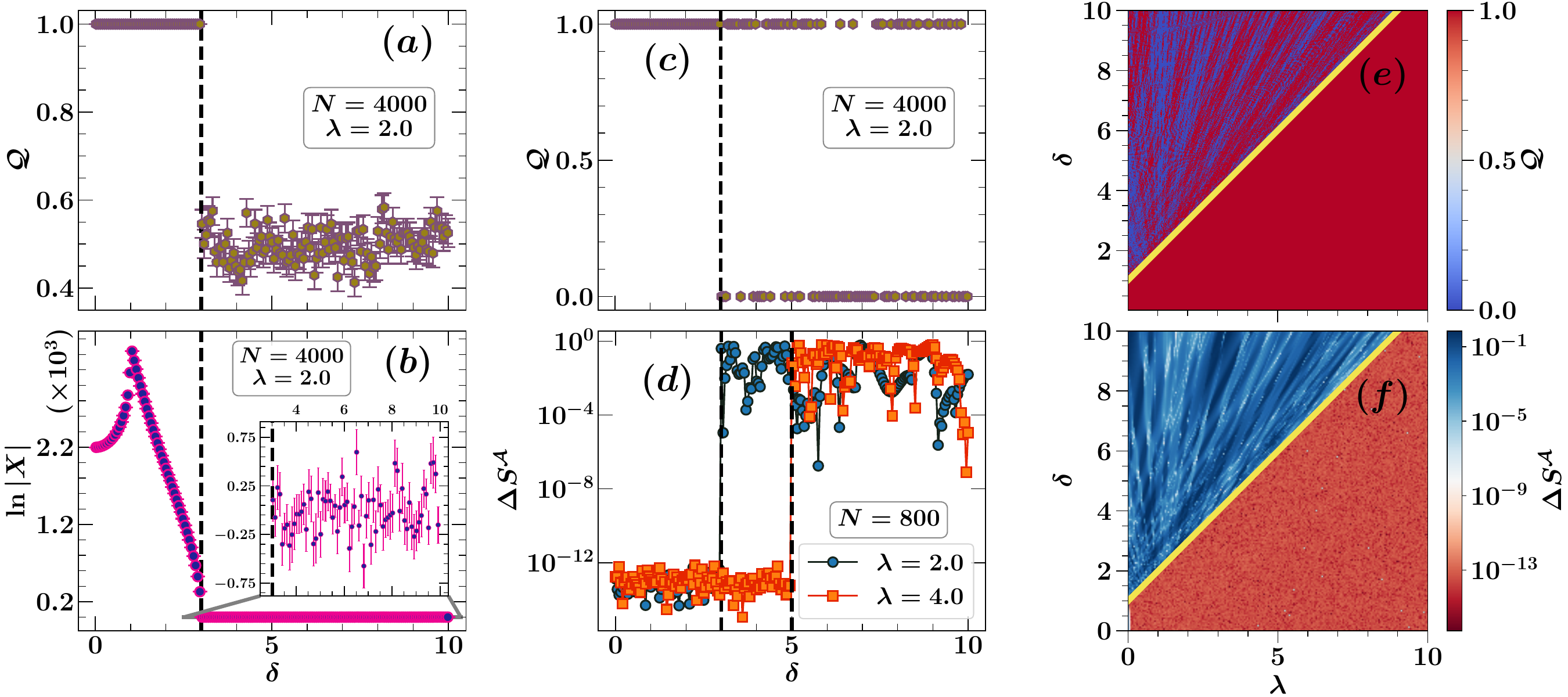}
    \caption{Characterization of the topological phase transition and
      comparative reliability of diagnostics for the quasiperiodic SSH
      model $ t = 1 $. Panels (a)-(d) show data for a fixed
      $\lambda=2.0$ as a function of disorder strength $\delta$,
      while panels (e) and (f) present full phase diagrams in the
      $ \lambda $-$ \delta $ plane for a single disorder
      realization. (a) Topological quantum number $\mathcal{Q}$
      ($N=4000$), averaged over 480 configurations. The system
      transitions from the topological phase ($ \mathcal{Q}=1 $) to
      the trivial phase ($ \mathcal{Q}=0 $) near the analytical
      phase boundary $ \delta = t + \lambda = 3.0 $ (black dashed
      line). In the trivial regime ($\delta > 3.0$), the average
      value of $\mathcal{Q}$ converges to approximately $0.5$
      instead of zero, indicating a failure to clearly distinguish the
      phases upon averaging. (b) The behavior of $\ln|X|$, the
      underlying quantity from which $\mathcal{Q}$ is derived (see
      Eqs.~(\ref{eq:lnX}) and
      (\ref{eq:topological-quantum-number})). In the topological
      phase, $\ln|X| > 0$, but in the trivial phase
      ($\delta > t+\lambda$), it fluctuates symmetrically around
      zero. The inset shows a zoomed-in view of these
      fluctuations. These fluctuations explain the averaged value of
      $ \mathcal{Q} \approx 0.5$ observed in (a), as $\mathcal{Q}$
      takes values of $1$ and $0$ with roughly equal probability
      when $\ln|X|$ changes sign. (c, d) Direct comparison of
      $ \mathcal{Q} $ and the entanglement entropy difference
      $ \Delta S^{\mathcal{A}} $ for a single disorder
      realization. (c) The value of $ \mathcal{Q} $ can be
      misleadingly equal to 1 in patches deep within the trivial phase
      ($ \delta > 3.0 $), indicating a lack of reliability for
      single configurations. (d) In contrast, the entanglement entropy
      difference
      $ \Delta S^{\mathcal{A}} = |S^{\mathcal{A}}_{\text{hf}} -
      S^{\mathcal{A}}_{\text{hf}+1}| $, calculated for a system of
      size $ N=800 $, provides a sharp distinction: it remains
      vanishingly small ($ \sim 10^{-12} $) in the topological phase
      and rises abruptly to finite values ($ > 10^{-5} $) in the
      trivial phase. This demonstrates that
      $ \Delta S^{\mathcal{A}} $ is a robust single-configuration
      diagnostic. (e, f) Phase diagrams from a single realization of
      quasiperiodic disorder, highlighting the contrast between the
      two diagnostics. (e) Phase diagram computed from
      $ \mathcal{Q} $ ($ N=4000 $). While the topological phase
      ($ \delta < t+\lambda $, red) is correctly identified,
      $ \mathcal{Q} $ spuriously yields a value of 1 in numerous
      patches within the trivial phase ($ \delta > t+\lambda $),
      leading to a false-positive identification of topology. (f)
      Phase diagram computed from $ \Delta S^{\mathcal{A}} $
      ($ N=800 $). The transition is unambiguous:
      $ \Delta S^{\mathcal{A}} $ is effectively zero (red) in the
      topological phase and becomes finite (blue) in the trivial
      phase, sharply following the analytical boundary
      $ \delta = t + \lambda $ (yellow line). These full phase
      diagrams confirm that the entanglement entropy difference is a
      superior and reliable metric for distinguishing topological from
      trivial phases, even without configuration averaging.}
    \label{fig:combined-plots}
  \end{figure*}

\section{Summary and Outlook}
\label{sec:Summary}
In this work, we have demonstrated that entanglement
entropy (EE) can serve as an effective probe for detecting
topological phase transitions in the Su-Schrieffer-Heeger
(SSH) model and its disordered variants. We study the clean SSH model as well as two extensions
incorporating quasiperiodic and random binary disorder in the hopping
terms. We show that the entanglement entropy of the many-body ground
state at half-filling, one particle above, and one particle below
half-filling remains identical in the topological regime, while it
differs in the trivial regime. This behaviour of EE can be understood
by studying the occupation number of the sites of the full system,
which reveals that the extra particle after half filling occupies the
edge sites and hence does not contribute to the entanglement entropy
of the subsystem which is deep inside the bulk.

We further show that edge localization near half filling alone is not
sufficient to establish topological character by explicitly analyzing
domain-wall configurations in the SSH chain. In particular, we
demonstrate that non-topological domain-wall states can also give rise
to vanishing entanglement-entropy differences. By systematically
tuning the chosen subsystem, we show that genuine topological
zero-energy states remain robust exhibiting vanishing
$\Delta S^{\mathcal{A}}_{\text{DW}}$ whereas trivially localized
finite energy states lose this property. This domain wall analysis
establishes a systematic protocol in which entanglement entropy
diagnoses localization near half filling as a necessary condition,
while the topological nature is confirmed through an additional check
of robustness against shrinking subsystem size.

\begin{figure}
    \centering
    \includegraphics[width=0.48\textwidth]{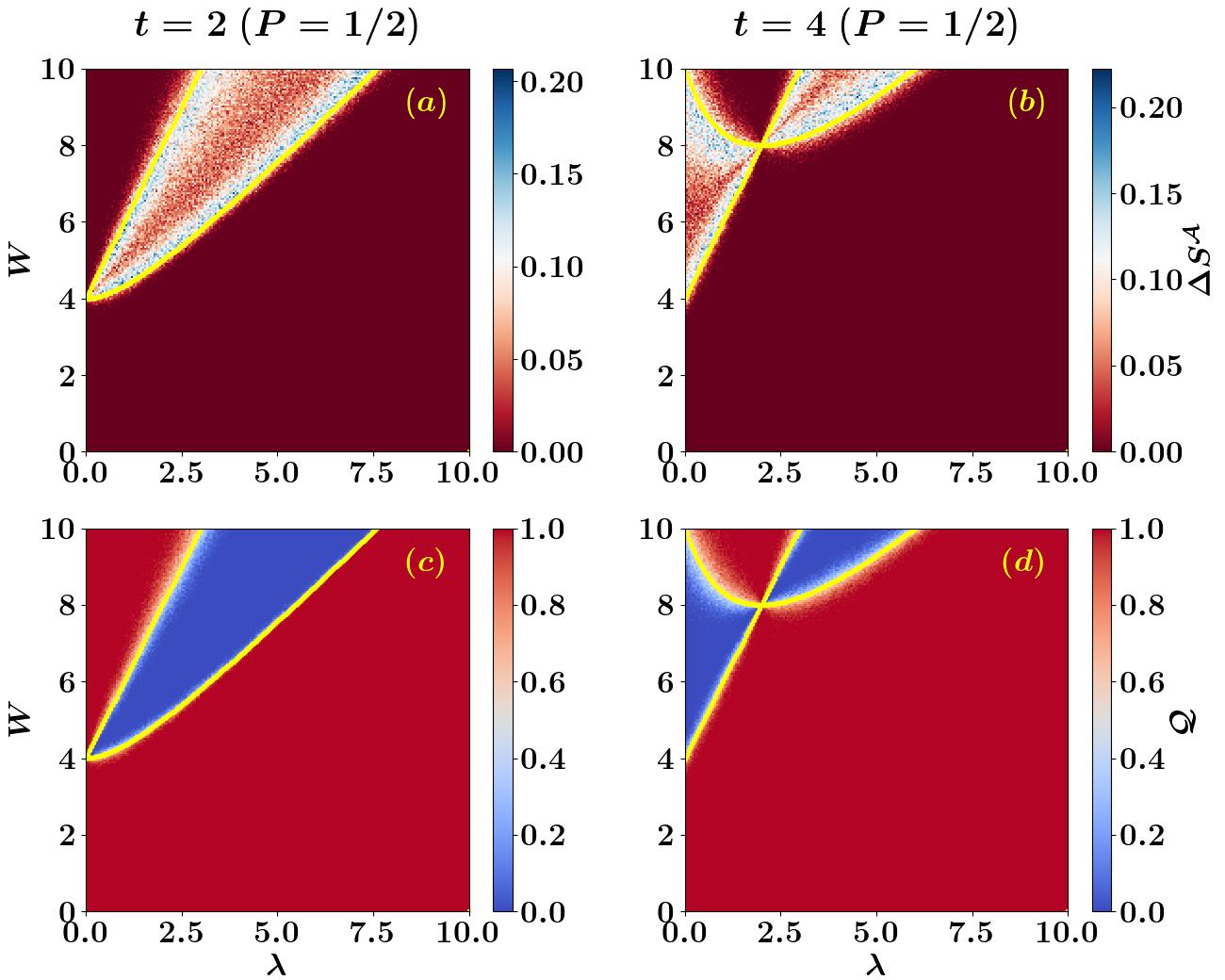}
    \caption{Phase diagrams for the SSH model with binary disorder
      with probability $P=1/2$. (a,b) Entanglement entropy difference
      $\Delta S^{\mathcal{A}} = |S^{\mathcal{A}}_{\text{hf}} -
      S^{\mathcal{A}}_{\text{hf}+1}|$ and (c,d) topological quantum
      number $\mathcal{Q}$, plotted as functions of binary disorder
      strength $W$ and dimerization $\lambda$. Left column (a,c):
      $t = 2$; right column (b,d): $t = 4$.
      $ \Delta S^{\mathcal{A}} $ is calculated at half filling with
      $ N_{\mathcal{A}} = 50 $. For all figures the system size is
      $ N = 400 $. For all the cases, the averaging is done over 100
      disorder realizations. Yellow curves denote analytical phase
      boundaries derived from Lyapunov exponents, separating
      topological (red) and trivial (blue) phases. The agreement
      between $\Delta S^{\mathcal{A}}$, $\mathcal{Q}$, and analytical
      boundaries confirms the robustness of our approach. Red (blue)
      regions denote $\mathcal{Q} = 1$ ($\mathcal{Q} = 0$) and
      $\Delta S^{\mathcal{A}} = 0$ ($\Delta S^{\mathcal{A}} > 0$),
      corresponding to topological and trivial phases respectively.}
    \label{appendix_P_1by2}
\end{figure}

\begin{figure}
    \centering
    \includegraphics[width=0.48\textwidth]{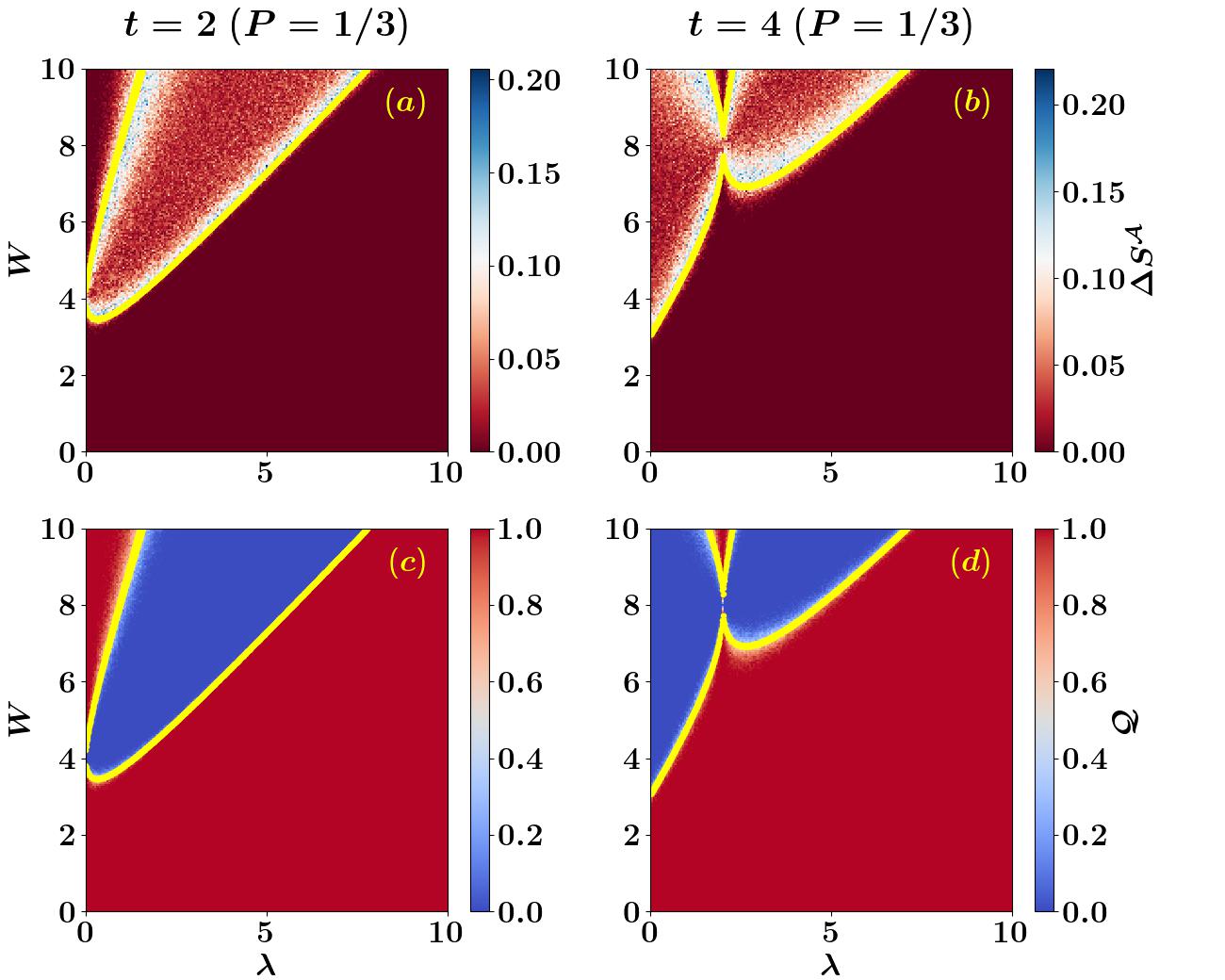}
    \caption{Phase diagrams for the SSH model with binary disorder
      with probability $P = 1/3$. The entanglement entropy difference
      $\Delta S^{\mathcal{A}}$ (a,b) and the topological quantum
      number $\mathcal{Q}$ (c,d) are shown as functions of disorder
      strength $W$ and dimerization $\lambda$. Results for two hopping
      strengths are presented: $t = 2$ (left column) and $t = 4$
      (right column). $ \Delta S^{\mathcal{A}} $ is calculated at
      half filling with $ N_{\mathcal{A}} = 50 $. For all figures
      the system size is $ N = 400 $. Data are averaged over 100
      disorder realizations. The analytically derived phase boundaries
      (yellow curves) excellently match with the numerical results,
      robustly distinguishing the topological ($\mathcal{Q}=1$,
      $\Delta S^{\mathcal{A}}=0$; red) and trivial ($\mathcal{Q}=0$,
      $\Delta S^{\mathcal{A}}>0$; blue) phases.}
    \label{P_1by3_DeltaQ_color}
\end{figure} 

In addition to numerical evidence, we develop a general analytic
framework based on the Lyapunov exponent (LE) to determine the
boundaries between topological and trivial phases. The LE is computed
using the transfer matrix method. This analytical framework is
straightforward and can be used to obtain the expression of boundaries
in systems with complex disorders also. We find excellent agreement
between our analytical predictions and numerical results across all
models studied, even in the presence of disorder. Finally, we validate
our conclusions using the topological quantum number $\mathcal{Q}$, a
known topological invariant, and find full consistency in the case of
clean SSH and random binary disorder models. However, in the presence
of quasiperiodic disorder in the hopping terms, $ \mathcal{Q} $
exhibits ambiguous behavior, particularly in the trivial regime. To
investigate this further, we compare $ \mathcal{Q} $ with the EE-based
diagnostic $ \Delta S^{\mathcal{A}} $, and find that the latter
vanishes in the topological phase even when a single disorder
realization is considered.

In this work, we establish a new quantity, $\Delta S^{\mathcal{A}}$,
which probes topological phase transitions in the 1D SSH model. While
1D systems host only two edge modes, higher-dimensional topological
insulators (e.g., 2D Chern insulators or 3D $\mathbb{Z}_2$ topological
insulators) exhibit richer edge/surface state structures. It is
therefore compelling to extend this entanglement-based diagnostic to
2D and 3D systems, where the interplay between bulk topology and
entanglement scaling remains incompletely explored. It is well known
that topological phases are accompanied by characteristic patterns of
long-range entanglement in the ground state~\cite{zeng2018}. By using
entanglement entropy (a quantum-information concept) to detect
topological phases (a many-body phenomenon), our work contributes to
the growing field of research of unifying quantum information science
and condensed matter physics. Finally, we note that the use of
entanglement spectrum to diagnose topological phases has also been
explored in earlier work by Fidkowski~\cite{PhysRevLett.104.130502};
establishing a bridge between Fidkowski's approach and our method is
desirable.

%In this work we have established a new quantity $\Delta S^A$ which can be used to as a probe of topological insulator in 1 dimensional model. 1D model have only two edge modes. In literature 2D systems with topological insulating states can be easily found. It will be interesting to test this new quantity in 2D systems as well as 3D systems where more than 2 edge modes exists. Since this is the era of unification of matter and quantum information, our work shows a direct relation of a quantum information concept, i.e. entanglement entropy, and a condensed matter concept, i.e. topological insulator which is a kind of quantum phase. Hence, our work can be one more step towards unifying information and matter.

\section*{Acknowledgments}
We thank Ivan M. Khaymovich for valuable discussions.  We are grateful to
the High Performance Computing (HPC) facility at IISER Bhopal, where
some calculations in this project were run. M.K is grateful to the
Council of Scientific and Industrial Research (CSIR), India, for his
PhD fellowship.

\appendix

\section{Topological quantum number}
\label{Quantum number}
To support our claims obtained using the entanglement
entropy, we use an established quantity, the topological quantum
number $\mathcal{Q}$, to distinguish a topological phase from a trivial
phase, which works even in the presence of disorder. The topological quantum number is defined as
\begin{equation}
  \mathcal{Q} = \nu(r),
\end{equation}
where $ r $ is the reflection matrix and $ \nu(r) $ denotes the
number of negative eigenvalues of $ r $~\cite{PhysRevB.83.155429,
  Zhang_2016}. A negative eigenvalue of $ r $ corresponds to a phase
shift of $ \pi $ in the reflected wave, indicating complete
reflection due to the presence of a zero-energy boundary
state~\cite{PhysRevLett.106.057001}. For $ N_C $ coupled SSH chains,
$ \nu(r) \in \{0, 1, 2, \dots, N_C\} $, and this corresponds to the
number of pairs of zero-energy edge modes.

For a single SSH chain, the reflection matrix $ r $ is a scalar and is given by
\begin{equation}
  r = \frac{1 - X^2}{1 + X^2},
  \label{r_exp}
\end{equation}
where
\begin{equation}
  X = (-1)^{N/2} \prod_{n=1}^{N/2} \frac{t_{2,n}}{t_{1,n}},
  \label{eq:lnX}
\end{equation}
with $ t_{1,n} $ and $ t_{2,n} $ being the intra-cell and inter-cell hoppings of the $ n^{\text{th}} $ unit cell, respectively. Hence, for a single SSH chain, the topological quantum number can be written as
\begin{equation}
  \mathcal{Q} = \frac{1}{2}(1 - \mathcal{Q}'),
  \label{eq:topological-quantum-number}
\end{equation}
where $ \mathcal{Q}' = \text{sign}(r) \in \{-1, 1\} $. Hence, in the
trivial phase, $ \mathcal{Q} = 0 $ (no edge modes); in the
topological phase, $ \mathcal{Q} = 1 $ due to one pair of
zero-energy edge modes.

To corroborate the results derived from $\Delta S^{\mathcal{A}}$, we
present phase diagrams for all three models in
Figs.~\ref{fig:Delta_S_color_plot}\flc{d-f}. For the cases of clean
SSH (Fig.~\ref{fig:Delta_S_color_plot}\flc{d}) and binary random
disorder (Fig.~\ref{fig:Delta_S_color_plot}\flc{f}), our results show
the expected behavior, i.e. $\mathcal{Q}=1$ in the topological regime
and $\mathcal{Q}=0$ in the trivial regime. The analytically calculated
boundaries, shown as yellow lines, match precisely with the numerical
results. These findings are fully consistent with the observations
from the entanglement entropy and further confirm that
$\Delta S^{\mathcal{A}}$ is an effective measure for distinguishing
topological and trivial phases.

For the case of quasiperiodic disorder,
Fig.~\ref{fig:Delta_S_color_plot}\flc{e} shows that
$\mathcal{Q} = 1$ throughout the entire topological regime. However,
in the trivial regime, $\mathcal{Q}$ deviates from the expected
behavior by acquiring nonzero values. Specifically, for
$\lambda=2$, when averaged over multiple values of $\phi$,
$\mathcal{Q}$ converges to approximately $0.5$ in the trivial
regime (see Fig.~\ref{fig:combined-plots}\flc{a}).  To understand this
average value of $\mathcal{Q} \approx 0.5$ better, we examine the
behavior of $\mathcal{Q'}=\text{sign}(1-X^2)=-\text{sign}(\log|X|)$,
which changes sign at the phase
transition. Figure~\ref{fig:combined-plots}\flc{b} shows that in the
topological regime, $ \log|X| > 0 $, while in the trivial regime,
$ \log|X| $ fluctuates around zero. This fluctuation is the origin
of the average $\mathcal{Q} \approx 0.5$. To gain deeper insight, we
next study $\mathcal{Q}$ for a single realization.

\subsection{Comparison between $\mathcal{Q}$ and $\Delta S^{\mathcal A}$}

Here, we compare the topological quantum number $\mathcal{Q}$ with
the entanglement-based quantity $\Delta S^{\mathcal{A}}$ for a
single realization of the quasiperiodic phase $\phi$. In
Fig.~\ref{fig:combined-plots}\flc{c}, we plot $\mathcal{Q}$ as a
function of $\delta$ for $\lambda = 2$. While $\mathcal{Q} = 1$
across the entire topological region ($\delta < t+\lambda$), it
fluctuates between 0 and 1 in the trivial regime
($\delta > t+\lambda$).  A similar behavior of $\mathcal{Q}$ is
visible in Fig.~\ref{fig:combined-plots}\flc{e} for all values of
$(\delta, \lambda)$ in the interval $[0, 10]$. Interestingly,
patches with $\mathcal{Q} = 1$ appear even within the trivial
region, which could falsely suggest topological insulating
behavior. This demonstrates that $\mathcal{Q}$ cannot sharply
identify the boundary between topological and trivial phases.  In
contrast, our proposed diagnostic, $\Delta S^{\mathcal{A}}$,
provides a sharp and reliable distinction between the two phases, as
shown in Figs.~\ref{fig:combined-plots}\flc{d,f}.

\section{Additional results for random binary disorder case}
\label{Appendix}
Here, we present our numerical results for the SSH model with random
binary disorder in the hopping amplitudes. We first show the plots for
probability $ P = 1/2 $ for various values of $ t $. The
topological phase is clearly distinguished from the trivial phase in
the $ W $--$ \lambda $ plane, as shown by the entanglement entropy
difference $ \Delta S^{\mathcal{A}} $ in
Figs.~\ref{appendix_P_1by2}\flc{a}, \ref{appendix_P_1by2}\flc{b}, and
by the topological invariant $ \mathcal{Q} $ in
Figs.~\ref{appendix_P_1by2}\flc{c}, \ref{appendix_P_1by2}\flc{d},
thereby confirming the consistency of our numerical technique.
Additionally, the yellow curves represent the analytically obtained
phase boundaries, which cleanly separate the topological and trivial
phases, thereby validating our analytical method.

Next, we examine the case of $ P = 1/3 $. In this scenario, the disorder $ \Delta_n $ is distributed as:
\begin{equation}
\Delta_n =
\begin{cases}
2, & \text{with probability } 1/3, \\
2 - W, & \text{with probability } 2/3.
\end{cases}
\label{Delta_value_P1by3}
\end{equation}
The analytical expressions for the topological phase boundaries are
obtained by substituting $ P = 1/3 $ into Eq.~\eqref{general_P1} and
Eq.~\eqref{general_P2}. We validate these boundaries numerically using
both the entanglement entropy difference $ \Delta S^{\mathcal{A}} $
and the topological invariant $ \mathcal{Q} $, as shown in
Fig.~\ref{P_1by3_DeltaQ_color}. The excellent agreement between the
analytical and numerical results further confirms the robustness of
our approach.

% add references to refs.bib
\bibliography{refs}

\end{document}